\documentclass[a4paper,twocolumn,11pt]{quantumarticle}
\pdfoutput=1

\usepackage[utf8]{inputenc}
\usepackage[english]{babel}
\usepackage[T1]{fontenc}
\usepackage{amsmath}
\usepackage{hyperref}

\usepackage{graphicx}
\usepackage{bm}
\usepackage{color}
\usepackage{enumerate}
\usepackage{bbold,soul}
\usepackage{lipsum}
\usepackage{array}
\usepackage{hyperref}
\usepackage{textgreek}
\usepackage[dvipsnames]{xcolor}
\usepackage{tabularx}
\usepackage{graphics,epsfig,subfigure}
\usepackage{wrapfig}
\usepackage{bbm}
\usepackage{mathrsfs}
\usepackage{cleveref}
\usepackage{amsmath}
\usepackage{booktabs}
\usepackage{placeins} 
\graphicspath{{images/}}

\usepackage{dcolumn}

\usepackage[normalem]{ulem}

\def\be{\begin{equation}}
\def\ee{\end{equation}}

\newcommand{\op}[1]{\hat{ #1}}                

\newcommand{\dd}{{\rm d}}

\newcommand{\ddt}{\Delta \tau}

\definecolor{nblue}{rgb}{0.06,0.3,0.73}
\definecolor{nblack}{rgb}{0,0,0}
\definecolor{nred}{rgb}{0.9,0.1,0.1}
\definecolor{nmagenta}{rgb}{0.7,0.0,0.3}
\definecolor{neditcolor}{rgb}{0.3,0.3,0.9}

\newcommand{\np}{N_p}
\newcommand{\ns}{N_s}
\newcommand{\nth}{N_\mathrm{th}}
\newcommand{\neff}{\check{N}_\mathrm{eff}}

\begin{document}

\title{Time-dependent multiparameter estimation for quantum experiments via online-offline sequential Monte-Carlo method}

\author{Chattamas Manoworakul}
\email{chattamasm@gmail.com}
\affiliation{Optical and Quantum Physics Laboratory, Department of Physics, Faculty of Science, Mahidol University, Bangkok, 10400, Thailand}
\orcid{0009-0006-3922-5161}

\author{Jason F. Ralph}
\affiliation{Department of Electrical Engineering and Electronics, University of Liverpool, Brownlow Hill, Liverpool L69 3GJ, United Kingdom}
\orcid{0000-0002-4946-9948}

\author{Nattaphong Wonglakhon}
\affiliation{Centre for Quantum Computation and Communication Technology (Australian Research Council), Centre for Quantum Dynamics, Griffith University, Yuggera Country, Brisbane, QLD 4111, Australia}
\orcid{0009-0000-4850-6121}

\author{Areeya Chantasri}
\email{areeya.chn@mahidol.ac.th}
\affiliation{Optical and Quantum Physics Laboratory, Department of Physics, Faculty of Science, Mahidol University, Bangkok, 10400, Thailand}
\affiliation{Centre for Quantum Computation and Communication Technology (Australian Research Council), Centre for Quantum Dynamics, Griffith University, Yuggera Country, Brisbane, QLD 4111, Australia}
\orcid{0000-0001-9775-536X}

\maketitle

\begin{abstract}
While typical online estimation methods can estimate the multiparameter dynamics of many systems, they may not be sufficient for a system with highly noisy measurement and rapid detection rate. In this paper, we create a hybrid estimation method by augmenting the sequential Monte Carlo (SMC) sampler, an online estimation method with an offline technique known as the time-batch estimation technique. By continuously monitoring the system, we may divide signals into batches and average them into an averaged trajectory. The system dynamics is then evolved with batch-averaged Kraus maps, for which we derive a highly efficient approximation. To facilitate the adoption of our algorithm, we present a modular derivation of the SMC methods and showcase our algorithm as an explicit example. We then implement our algorithm on the measurement signals obtained from superconducting-qubit experiments under two types of measurement setting: a fluorescence measurement and a dispersive $z$-measurement. The algorithm's hyperparameter values are chosen from independent numerical simulation, while the accuracy of our estimation is validated by a signal reconstruction method. Our results show that, for the fluorescence case, our algorithm can estimate the system's parameters better than the standard calibration method, and, for the dispersive case, our estimation is capable of finding an unexpected jump in parameter values that the standard calibration method could not find.
\end{abstract}
\maketitle


\section{Introduction}

Reliable estimation of multiparameter dynamics for quantum systems is a prerequisite for building applications in quantum technology, particularly in experimental realizations, where the physical parameters may deviate over time. To characterize such parameters, quantum devices must be continuously monitored so that their information can be extracted over time. Parameter estimation based on continuous quantum measurement has been extensively investigated over the past decades, using different techniques ranging from Bayesian estimation~\cite{Chase2009, Ralph2011, Six2015, kiilerich2016, Ralph2017}, maximum-likelihood estimation~\cite{Cortez2017, Clausen2024}, Markov-chain Monte-Carlo estimation~\cite{Gammelmark2013}, a regression with correlation functions~\cite{Guilmin2024}, to using the recent machine learning models~\cite{Genois2021, Rinaldi2024, Tucker2024}.
Moreover, estimation protocols can also be adaptive or optimized with variable controls, e.g., combining with maximum-likelihood estimation~\cite{Radaelli2024} or Bayesian estimation~\cite{Godley2023, Pitchford2024,Bel2026}. 

From a time-dependent perspective and how observation data are processed, previous investigations suggest that parameter estimation methods can be broadly categorized into two regimes: \emph{online} and \emph{offline}. Online estimation algorithms process observed data as they arrive; they aim to provide (close to) real-time estimation or time-refined dynamics of the system's parameters using recursive estimation. 
One limitation of online estimation is a constraint on its estimation time. For an algorithm to respond to the system dynamics, its estimation time must be comparable to the detection timescale i.e., the algorithm must run at the timescale comparable to the data-acquisition time with low latency.
An example of an online estimation algorithm for quantum systems is Clausen et al. (2024)~\cite{Clausen2024}.
On the other hand, offline estimation processes the data at once, without a time constraint. Their algorithms can be computationally complex to achieve estimation results with the best accuracy possible. 
However, with offline estimation, one cannot iteratively update the estimates with newly available data, making them unsuitable for real-time estimation. An example of an offline estimation method for quantum systems is Gammelmark and Mølmer (2013)~\cite{Gammelmark2013}. In reality, however, the distinction between online and offline methods is somewhat arbitrary. Online estimation may benefit from offline preprocessing techniques, and offline estimation may benefit from online recursion technique.s A fusion of the two methods enables a more specialized algorithm for some specific scenarios.

For example, consider a realistic quantum experiment, where environmental interactions cause the system's parameters to drift. To track parameter dynamics, one would need a real-time (online) parameter estimation with a detection regime which can observe the system with fast detection rate, low latency, and high signal-to-noise ratio. These conditions are usually unfeasible in real experiments.
However, the quality of parameter tracking are not purely determined by the detection timescale. It is determined by roughly two factors: the detection timescale and the parameter-drift timescale. Typically, the former is faster than the latter. An obvious strategy is to aim for an algorithm to be \emph{online} with respect to the parameter-drift timescale, allowing it to be \emph{offline} with respect to the detection timescale. A way to accomplish this is to combine the online and offline estimation methods into a \emph{hybrid} parameter estimation method. By operating in the timescale between the measurement resolution and the parameter drift, the hybrid method can reduce measurement noise while outputting online estimation results.

Our choice of hybrid estimation algorithm originates from the Sequential Monte Carlo (SMC) methods. One of the earliest quantum applications of the methods is the work by Chase and Geremia in 2009~\cite{Chase2009}, where they developed a quantum particle filter for continuous-valued parameters. In 2015, Six et al.~\cite{Six2015} demonstrated the stability of the particle filter by estimating a single parameter from experimental data. A similar method, the Sequential Monte Carlo sampler (SMC sampler), was implemented in Ralph et al. (2017)~\cite{Ralph2017} for a more general type of problems. While the aforementioned methods were online, our proposed algorithm is a hybrid estimation method: the SMC sampler augmented with the batch estimation technique~\cite{Chopin2002}. The batch estimation technique divides the time-series dataset into batches, limiting the complexity of the algorithm while averaging multiple noisy trajectories into a less noisy one. This allows the algorithm to track the dynamics of the parameters and iteratively refine their estimated values using prior knowledge. Since the SMC methods are modular, any practitioner can easily adapt the derived formulae for their use cases. 
However, as the SMC methods involve recursive Bayesian statistics and Monte Carlo sampling, its technical details can be inaccessible to non-specialists. We thus provide in this work a practical formulation of the SMC methods, specifically the particle filters and the SMC sampler. Our goal is to give an intuitive construction of the methods, with an explicitly derived example of the SMC sampler augmented with the batch estimation technique. 
We note that excellent in-depth reviews of the two methods already exist~\cite{Arulampalam2002, Cappe2007, Dai2022}. For those interested in the convergence proofs, see Crisan and Doucet (2002)~\cite{Crisan2002} for the particle filters and Del Moral et al.~(2006)~\cite{Del2006} for the SMC sampler.

Aside from the hybrid SMC algorithm, our paper contains two more significant contributions. Firstly, we account for the noise reduction of the batch estimation technique~\cite{Chopin2002} by introducing a procedure for trajectory-averaged dynamics in the form of a Kraus map. Our approximated map, which is derived in Appendix~\ref{app:fullMap}, is much more efficient than the exact form. Secondly, we implement our derived Kraus map and augmented SMC algorithm on experimental datasets from two superconducting experiments: a transmon qubit under a continuous fluorescence measurement~\cite{Naghiloo2016, Naghiloo2017} and a dispersive $z$-measurement~\cite{Weber2014}. The reasons for choosing these types of experiments are twofold: not only are the superconducting qubit experiments currently the most practical platforms for quantum computing~\cite{Kjaergaard2020, Jiang2025}, but they also exhibit two distinct timescales, i.e., the faster detection time (with low latency and signal-to-noise ratio) and the slower parameter drift; these properties make them a suitable scenario for our hybrid estimation method.
Additionally, we have used simulated datasets to determine appropriate hyperparameter values of our algorithm. Then, we compare the results from our algorithm to parameters obtained from standard calibration methods. Using the estimated parameters to reconstruct the signals, we have shown that our algorithm has comparable accuracy to the standard calibration method. Moreover, for the dispersive $z$-measurement case, our algorithm captures a sudden jump in the dynamics that the calibration method did not find.

The structure of our paper is as followed: Section~\ref{sec-smc} develops the SMC methods, specifically the particle filters and the SMC sampler. The first four subsections formulate the general theory of the two methods while the last two subsections are the explicit examples of our algorithm. In Section~\ref{sec:qTraj}, the quantum trajectory formalism is introduced as the model for our quantum dynamics. Following the idea of batch estimation, we give a simple trajectory-averaged Kraus map, an approximate form of which we derive in Appendix~\ref{app:fullMap}. We implement our algorithm in Section~\ref{sec-implement}, starting with an explicit protocol detailed in the first subsection, followed by two subsections detailing the estimation results with the two experimental datasets. Lastly, Section~\ref{sec:conC} concludes our finding and provides future directions.

\section{Sequential Monte Carlo Method}\label{sec-smc}

In this section, we develop two SMC methods for parameter estimation problems. Section~\ref{subsec:PS} presents a problem statement and its formal solution. Since analytical solution is often unfeasible, Section~\ref{subsec:IS} showcases the importance sampling method, a numerical method to solve highly-complex integrals. We can improve its efficiency by taking advantage of the dynamical structure of the problem. Section~\ref{subsec:SIS} introduces a sequential version of the importance sampling and discusses the resampling step, which alleviates the particle degeneracy problem~\cite{Arulampalam2002}. Section~\ref{subsec:SMCS} develops an alternative to the sequential importance sampling. In Section~\ref{subsec:combinedSMC}, the two algorithms are combined to better accommodate our problem of interest. Finally, Section~\ref{subsec:longSMCS} augments the final algorithm with the batch estimation technique, resulting in the algorithm we shall use throughout the paper.

We note that our work focus only on practicality and forego much of the theoretical details. For readers interested in a boarder treatment of the SMC methods, see Doucet et al. (2001)~\cite{Doucet2001}.

\subsection{Problem statement: estimation of time-dependent parameters}\label{subsec:PS}

A typical parameter estimation problem involves determining a set of unknown system parameters $\mathbf{x}_t \in \mathcal{X}$ where $\mathcal{X}$ is the parameter space and the subscript $t$ is the time step. This can be done using the information encoded in a set of measurement results $\mathbf{y}_t \in \mathcal{Y}$ where $\mathcal{Y}$ is the observation space. Specifically, we are interested in the time evolution of the parameters $\mathbf{x}_{0:t} \equiv \{\mathbf{x}_0, \mathbf{x}_1, \ldots, \mathbf{x}_t\}$ given a series of measurements $\mathbf{y}_{1:t}$ where $\mathbf{x}_0$ is the set of initial parameters. We assume the system's dynamics to be Markovian, though we note that the result readily generalizes to the non-Markovian case. Therefore, the parameters evolve through the transition distribution $\mathbf{x}_t \sim g(\mathbf{x}_t|\mathbf{x}_{t-1})$, with the measurement data through the observation distribution $\mathbf{y}_t \sim f(\mathbf{y}_t|\mathbf{x}_t)$.

As with most parameter estimation problems, we wish to find the posterior distribution $p(\mathbf{x}_{0:t}|\mathbf{y}_{1:t})$ and its associated features, such as the marginal distribution $p(\mathbf{x}_t|\mathbf{y}_{1:t})$ and an expectation of an arbitrary function $h(\mathbf{x}_{0:t})$,
\begin{equation} \label{eqn:goalSMC}
    \mathbbm{E}_{\mathbf{x}_{0:t}|\mathbf{y}_{1:t}}[h(\mathbf{x}_{0:t})] \equiv \int h(\mathbf{x}_{0:t})p(\mathbf{x}_{0:t}|\mathbf{y}_{1:t}) \, {\rm d}\mathbf{x}_{0:t}.
\end{equation}

Formally, the posterior distribution may be obtained recursively through Bayes's theorem,
\begin{equation} \label{eqn:BayesPost}
    p(\mathbf{x}_{0:t}|\mathbf{y}_{1:t}) = p(\mathbf{x}_{0:t-1}|\mathbf{y}_{1:t-1}) \frac{f(\mathbf{y}_{t}|\mathbf{x}_{t})g(\mathbf{x}_{t}|\mathbf{x}_{t-1})}{p(\mathbf{y}_{t}|\mathbf{y}_{1:t-1})},
\end{equation}
where $p(\mathbf{y}_{t}|\mathbf{y}_{1:t-1}) = \int f(\mathbf{y}_t|\mathbf{x}_t)p(\mathbf{x}_t|\mathbf{y}_{1:t-1})\, {\rm d}\mathbf{x}_t$. Alternatively, if only the marginal distribution $p(\mathbf{x}_t|\mathbf{y}_{1:t})$ is of interests, we may compute it through an update-prediction recursion~\cite{Doucet2001},
\begin{align} \label{eqn:update}
    p(\mathbf{x}_t|\mathbf{y}_{1:t-1}) =& \int \!\! g(\mathbf{x}_t|\mathbf{x}_{t-1}) p(\mathbf{x}_{t-1}|\mathbf{y}_{1:t-1}) \, {\rm d}\mathbf{x}_{t-1}, \\ \label{eqn:prediction}
    p(\mathbf{x}_t|\mathbf{y}_{1:t}) =& \frac{f(\mathbf{y}_t|\mathbf{x}_t) p(\mathbf{x}_t|\mathbf{y}_{1:t-1})}{\int f(\mathbf{y}_t|\mathbf{x}_t) p(\mathbf{x}_t|\mathbf{y}_{1:t-1}) \, {\rm d}\mathbf{x}_t}.
\end{align}
Despite being recursive, these formulas do not permit tractable analytical solutions except for a few special cases, as they require the evaluation of highly complex distribution: $p(\mathbf{y}_{t}|\mathbf{y}_{1:t-1}) = \int f(\mathbf{y}_t|\mathbf{x}_t)p(\mathbf{x}_t|\mathbf{y}_{1:t-1})\, {\rm d}\mathbf{x}_t$. When analytical solutions are intractable, we turn to Monte Carlo methods for numerical approximation.

\subsection{Monte Carlo integration and \\ Importance sampling} \label{subsec:IS}
There is a simple probabilistic numerical method for estimating the expectation $\mathbbm{E}_{p(\mathbf{x}_{0:t}|\mathbf{y}_{1:t})}[h(\mathbf{x}_{0:t})]$ in Eq.~\eqref{eqn:goalSMC}, which is known as \textit{Monte Carlo integration}. Instead of evaluating the entire integral's domain, we sample $\np$ points (also known as particles) from the posterior distribution $p(\mathbf{x}_{0:t}|\mathbf{y}_{1:t})$ and compute an approximation of the expectation,
\begin{equation}
    \mathbbm{E}_{\mathbf{x}_{0:t}|\mathbf{y}_{1:t}}[h(\mathbf{x}_{0:t})] \approx \frac{1}{\np} \sum_i^{\np} h(\mathbf{x}_{0:t}^{(i)}),
\end{equation}
where the summation is over all the points sampled from the posterior distribution  $\mathbf{x}_{0:t}^{(i)} \sim p(\mathbf{x}_{0:t}|\mathbf{y}_{1:t})$. The estimation accuracy increases as the size of sample points $N$ gets large.

Oftentimes it is difficult to efficiently sample from the posterior distribution $p(\mathbf{x}_{0:t}|\mathbf{y}_{1:t})$ due to its high dimensionality. A more effective method known as the \textit{importance sampling} can be used to sample $\mathbf{x}_{0:t}$ from an arbitrary distribution with more flexible properties~\cite{Tokdar2010}. This can be done by introducing a \emph{proposal distribution}, $q(\mathbf{x}_{0:t}|\mathbf{y}_{0:t})$, whose support includes the support of $p(\mathbf{x}_{0:t}|\mathbf{y}_{1:t})$, i.e. any point $\mathbf{x}_{0:t}$ that can be sampled from $p(\mathbf{x}_{0:t}|\mathbf{y}_{0:t})$ can be sampled from $q(\mathbf{x}_{0:t}|\mathbf{y}_{1:t})$. Then, we rewrite the expectation value in Eq.~\eqref{eqn:goalSMC} as,
\begin{equation}\label{eq-impsampling}
    \mathbbm{E}_{\mathbf{x}_{0:t}|\mathbf{y}_{1:t}}[h(\mathbf{x}_{0:t})] = \frac{\int h(\mathbf{x}_{0:t}) w(\mathbf{x}_{0:t}) q(\mathbf{x}_{0:t}|\mathbf{y}_{1:t}) \, {\rm d}\mathbf{x}_{0:t}}{\int w(\mathbf{x}_{0:t}) q(\mathbf{x}_{0:t}|\mathbf{y}_{1:t}) \,{\rm d}\mathbf{x}_{0:t}}
\end{equation}
where we have introduced an unnormalized importance weight defined as
\begin{align}\label{eq-weight}
    w(\mathbf{x}_{0:t}) \equiv \frac{p(\mathbf{x}_{0:t}|\mathbf{y}_{1:t})}{q(\mathbf{x}_{0:t}|\mathbf{y}_{1:t})}.
\end{align}
With the Monte Carlo numerical method, the integral in Eq.~\eqref{eq-impsampling} is approximated by
\begin{equation} \label{eqn:empMean}
    \mathbbm{E}_{\mathbf{x}_{0:t}|\mathbf{y}_{1:t}}[h(\mathbf{x}_{0:t})] \approx \sum_i^{\np} h(\mathbf{x}_{0:t}^{(i)}) \tilde{w}_t^{(i)},
\end{equation}
where $\tilde{w}_t^{(i)} \equiv w(\mathbf{x}_{0:t}^{(i)})/\sum_j^N w(\mathbf{x}_{0:t}^{(j)})$ and the summation here is over all the points sampled from the proposal distribution $\mathbf{x}_{0:t}^{(i)} \sim q(\mathbf{x}_{0:t}|\mathbf{y}_{1:t})$. 

The importance sampling is still not efficient for recursive parameter estimation. Whenever a new data point $\mathbf{y}_{t+1}$ is acquired, the new importance weight $\tilde{w}_{t+1}^{(i)}$ must be calculated via the new posterior and proposal distributions from Eq.~\eqref{eq-weight}. This means that the complexity of the calculation will increase with time.

\subsection{Sequential Importance Sampling (SIS)} \label{subsec:SIS}

To alleviate the inefficiency of the importance sampling, we may choose a family of proposal distribution $q(\mathbf{x}_{0:t}|\mathbf{y}_{1:t})$ with a sequential property to circumvent the need to resample the whole high-dimensional space of $\mathbf{x}_{0:t}$ whenever new measurement data is acquired. This specialized version of the importance sampling is called \textit{sequential importance sampling} (SIS), also known as the \textit{particle filter}. The technique belongs to the wider class of methods known as \textit{sequential Monte Carlo} (SMC) methods, which aim to reduce computational cost for sequential estimation problems.

In sequential importance sampling, the proposal distribution is chosen such that it admits the marginal $q(\mathbf{x}_t|\mathbf{x}_{0:t-1}, \mathbf{y}_{1:t})$,
\begin{equation}\label{eq-sis}
    q(\mathbf{x}_{0:t}|\mathbf{y}_{1:t}) = q(\mathbf{x}_t|\mathbf{x}_{0:t-1}, \mathbf{y}_{1:t}) q(\mathbf{x}_{0:t-1}|\mathbf{y}_{1:t-1}).
\end{equation}
By combining the importance weight $w(\mathbf{x}_{0:t})$ in Eq.~\eqref{eq-weight} and the posterior distribution in Eq.~\eqref{eqn:BayesPost}, we can utilize the required property from Eq.~\eqref{eq-sis} to derive a recursion relation for the normalized weight of the $i$-th particle as 
\begin{equation} \label{eqn:seqImpWeight}
    \tilde{w}_t^{(i)} \propto \tilde{w}_{t-1}^{(i)}
    \frac{f(\mathbf{y}_{t}|\mathbf{x}_{t}^{(i)})g(\mathbf{x}_{t}^{(i)}|\mathbf{x}_{t-1}^{(i)})}{q(\mathbf{x}_t^{(i)}|\mathbf{x}_{0:t-1}^{(i)}, \mathbf{y}_{1:t})},
\end{equation}
where the proportional sign is for the normalization~\cite{Doucet2001}. Eq.~\eqref{eqn:seqImpWeight} is a simple weight-update equation for each individual particle. For each time step $t$, the new parameter values of each particle $\mathbf{x}_t^{(i)}$ is drawn from the marginal $\mathbf{x}_t^{(i)} \sim  q(\mathbf{x}_t|\mathbf{x}_{0:t-1}, \mathbf{y}_{1:t})$ and used to update the weights using the recursion from Eq.~\eqref{eqn:seqImpWeight}.


Despite its efficiency, SIS suffers from a different problem than the importance sampling. For $t\gg 1$, most particles will have negligible weights due to how they are sampled.
This problem is known as particle degeneracy~\cite{Doucet2000}. We can alleviate it by introducing additional resampling steps
to decrease the number of low-weight particles. For example, the bootstrap filter resamples particles using a smoothing kernel of the empirical multinomial distribution, resulting in the selection-move step where the original particles are resampled and perturb to better approximate the posterior.
For a recent review of resampling algorithms, see Kuptametee and Aunsri (2022)~\cite{Kuptametee2022}.

\subsection{Sequential Monte Carlo (SMC) sampler}\label{subsec:SMCS}
 

Let us consider two other cases of our problem statement: one where we cannot justify the form of the transition distribution $g(\mathbf{x}_{t}|\mathbf{x}_{t-1})$, and another where the parameters $\mathbf{x}_t$ are static. For the former, we cannot compute the weight-update Eq.~\eqref{eqn:seqImpWeight} without the transition distribution. For the latter, the estimation from SIS may not converge to the true values, as the algorithm's proof of convergence relies on the noise processes of the transition distribution to explore the parameter space efficiently~\cite{Crisan2002,Maskell2012}. For these two cases, another SMC method, known as \textit{sequential Monte Carlo sampler} (SMC sampler), becomes more appropriate.

Our formulation of the SMC sampler is based on the development in Green and Maskell (2017)~\cite{Green2017}. To estimate the unknown parameters $\mathbf{x}_{0:t}$, we introduce the target distributions $\{\pi(\mathbf{x}_0), \pi(\mathbf{x}_1), \ldots, \pi(\mathbf{x}_t) \}$ whose joint distribution is $\pi(\mathbf{x}_{0:t})$. We note that the target distributions can be conditioned on observed data, e.g., $\pi(\mathbf{x}_t) = p(\mathbf{x}_t|\mathbf{y}_{1:t})$. To construct a weight-update recursion similar to Eq.~\eqref{eqn:seqImpWeight}, we impose a sequential structure to the joint distribution $\pi(\mathbf{x}_{0:t})$ by defining a time-reversal kernel $l(\mathbf{x}_{i-1}|\mathbf{x}_i)$ such that
\begin{equation}\label{eq-pireverse}
    \pi(\mathbf{x}_{0:t}) = \pi(\mathbf{x}_t)\prod_{i=1}^t l(\mathbf{x}_{i-1}|\mathbf{x}_i).
\end{equation}
Next, we construct a proposal distribution with the (time-forward) sequential structure $q(\mathbf{x}_{0:t}) = q(\mathbf{x}_t|\mathbf{x}_{t-1}) q(\mathbf{x}_{0:t-1})$, similar to Eq.~\eqref{eq-sis} of Section~\ref{subsec:SIS}, and define a weight $w(\mathbf{x}_{0:t}) = \pi(\mathbf{x}_{0:t})/q(\mathbf{x}_{0:t})$. Taking the ratio of the weights to obtain the new weight-update recursion,
\begin{equation} \label{eqn:weightSMCsampler}
    \tilde{w}_t^{(i)} \propto \tilde{w}_{t-1}^{(i)} \frac{\pi(\mathbf{x}_t^{(i)}) }{\pi(\mathbf{x}_{t-1}^{(i)})} \frac{l(\mathbf{x}_{t-1}^{(i)}|\mathbf{x}_t^{(i)})}{q(\mathbf{x}_t^{(i)}|\mathbf{x}_{t-1}^{(i)})},
\end{equation}
where $\tilde{w}_t^{(i)}$ is the normalized importance weight of the $i$-th particle. We note here that the target distributions as well as the proposal distributions can also be conditioned on the measurement readouts $\mathbf{y}_{1:t}$ as per the earlier example.

It is clear from Eq.~\eqref{eqn:weightSMCsampler} that the SMC sampler is more general than SIS; it does not require the transition distribution. Its proof of convergence does not require the noise processes~\cite{Del2006}, and its calculation ensures equal weighting for all measurement results whether they are from the past or the present~\cite{Maskell2012}.
However, it is important to note that the SMC sampler's recursion Eq.~\eqref{eqn:weightSMCsampler} does not offer the same efficiency as Eq. \eqref{eqn:seqImpWeight}. The target distribution $\pi(\mathbf{x}_t)$ does not have a recursive update formula. It must be calculated separately for each time step. Using the example of $\pi(\mathbf{x}_t) = p(\mathbf{x}_t|\mathbf{y}_{1:t})$, we can see that the complexity of the weight-update recursion scales with time. And like SIS, the SMC sampler requires a resampling step to combat particle degeneracy.

\subsection{Combined SMC methods for time-dependent parameter estimation} \label{subsec:combinedSMC}

Two distinct SMC methods have been developed in Sections \ref{subsec:SIS} and \ref{subsec:SMCS}. They have their advantages and disadvantages. SIS is more efficient but requires the transition distribution, while SMC sampler is more general but less efficient. In this section, we combine the two algorithms to alleviate their shortcomings. For our problem of interest, we do not know the form of the transition distribution, only that it is constant or slow-varying. For very large datasets, running the SMC sampler as the main algorithm will incur a huge computational cost. Therefore, we choose the SIS with zero-noise (static) model as the main algorithm and the SMC sampler as the resampling step to ensure convergence of the estimation. We note that this algorithm is used in Ralph et al. (2017) \cite{Ralph2017}. It is adapted from Green and Maskell (2017) \cite{Green2017}.

\emph{Main algorithm}: The main algorithm is SIS with a zero-noise transition $g(\mathbf{x}_{t}|\mathbf{x}_{t-1}) = \delta(\mathbf{x}_t - \mathbf{x}_{t-1})$. For simplicity's sake, we choose $q(\mathbf{x}_t|\mathbf{x}_{0:t-1}, \mathbf{y}_{1:t}) = g(\mathbf{x}_t|\mathbf{x}_{t-1})$. Substituting these in Eq.~\eqref{eqn:seqImpWeight}, we obtain the weight-update recursion,
\begin{equation} \label{eqn:ourWeightUpdate}
    \tilde{w}_t^{(i)} \propto  f(\mathbf{y}_t|\mathbf{x}_t^{(i)})\tilde{w}_{t-1}^{(i)}.
\end{equation}

\emph{Resampling step}: The SMC sampler is employed for the resampling step. For a measure of particle degeneracy, we have chosen the effective sample size~\cite{Bergman1999}. Its estimator is 
\begin{align}
    \neff = 1/\sum_j(\tilde{w}^{(j)}_t)^2.
\end{align}
The resampling step is initiated if $\neff < \nth$ where $\nth$ is a threshold value. There are two parts in the resampling step: selection and move steps. Assuming the resampling step is initialized at time step $k$, the selection step reselects \emph{pre-resampling} parameters $\mathbf{x}_{k}$ to get rid of low-weight particles while the move step turns them into \emph{post-resampling} parameters $\tilde{\mathbf{x}}_{k}$ through a proposal distribution $q(\tilde{\mathbf{x}}_k|\mathbf{x}_k)$. The selection process is done by sampling from the initial parameter set ${\mathbf{x}}_k^{(i)}$ based on the current weights, i.e., the set of $N$ particles $\{ {\mathbf{x}}_k^{(1)}, {\mathbf{x}}_k^{(2)},..., {\mathbf{x}}_k^{(N)}\}$ are re-selected using the multinomial distribution of the same particle set with the normalized weights $\{ \tilde{w}_k^{(1)}, \tilde{w}_k^{(2)},...,\tilde{w}_k^{(N)}\}$. Systematic resampling~\cite{Kitagawa1996} is used to facilitate the process. Particles with significant weights will be selected more often than particles with low weights. The new weights become $\tilde{w}_\mathrm{pre}^{(i)} \equiv 1/\np$.

The move step introduces variety to the parameter values by resampling the particles and updating their weights using the SMC sampler protocol. The resampling process is done by selecting the post-resampling particles $\tilde{\mathbf{x}}_k^{(i)}$ from a Gaussian smoothing kernel~\cite{Liu2001} with a defensive strategy~\cite{Hesterberg1995}. We define the empirical covariance,
\begin{align}
    \Sigma_k \equiv \mathbbm{E}[(\mathbf{x}_k - \mathbbm{E}[\mathbf{x}_k])(\mathbf{x}_k - \mathbbm{E}[\mathbf{x}_k])^\mathrm{T}],
\end{align}
where $\mathbbm{E}[\bullet]$ is an expectation over the particles $\mathbf{x}^{(i)}_k$ and $\bullet^\mathrm{T}$ is the matrix transpose. Additionally, we implement the defensive strategy by defining the ratio parameters $p$ and $q$ where $0<p,q<1$ such that $p\np$ out of $\np$ particles are resampled with the covariance $q\Sigma_k$ while the remaining $(1-p)\np$ particles are resampled with the covariance $\Sigma_k$.

The weights of the post-resampling particles are updated according to Eq.~\eqref{eqn:weightSMCsampler}. We take the pre- and post-resampling target distributions to be $\pi(\mathbf{x}_k) = p(\mathbf{x}_k|\mathbf{y}_{1:t})$ and $\pi(\tilde{\mathbf{x}}_k) = p(\tilde{\mathbf{x}}_k|\mathbf{y}_{1:t})$ respectively. The proposal distribution $q(\tilde{\mathbf{x}}_k|\mathbf{x}_{k})$ is the Gaussian smoothing kernel described in the move step. Due to the symmetry of the smoothing kernel, the time-reversal kernel can be chosen to be equal to the proposal $ l(\mathbf{x}_{k}|\tilde{\mathbf{x}}_k) = q(\tilde{\mathbf{x}}_k|\mathbf{x}_{k})$. Therefore, the weight-update Eq.~\eqref{eqn:weightSMCsampler} can be simplified to
\begin{equation}
\label{eqn:weightUpdateResampling}
    \tilde{w}_\mathrm{post}^{(i)} \propto \tilde{w}^{(i)}_\mathrm{pre} \frac{p(\tilde{\mathbf{x}}_k^{(i)}|\mathbf{y}_{1:k})}{p({\mathbf{x}}_k^{(i)}|\mathbf{y}_{1:k})} \propto \frac{\prod_{r=1}^k f(\mathbf{y}_r|\tilde{\mathbf{x}}_k^{(i)})}{\prod_{r=1}^k f(\mathbf{y}_r|\mathbf{x}_k^{(i)})},
\end{equation}
where, for the last proportional sign, we have used $\tilde{w}^{(i)}_\mathrm{pre} = 1/\np$ and the approximation of the priors $p(\mathbf{x}_k) \approx p(\tilde{\mathbf{x}}_k)$ since the proposal distribution is a smoothing kernel; the two empirical priors approximate the same distribution. We note that the resampling step can be repeated until the effective sample size $\neff$ goes above the threshold $\nth$.

\subsection{Tracking long-term parameter drift} \label{subsec:longSMCS}

Although the algorithm in Section~\ref{subsec:combinedSMC} is more efficient than a naive SMC sampler, its computational cost scales with time. For parameter tracking of a large timescale $k\gg1$, Eq.~\eqref{eqn:weightUpdateResampling} will require a huge amount of operations each time the resampling step is initialized. If the system of interest has rapid and noisy data detection rates, the algorithm may not be fast or accurate enough. This section fixes the issue with the technique known as \emph{batch estimation}, introduced in Chopin (2002)~\cite{Chopin2002}. Instead of one continuous estimation, the measurement data are partitioned into batches, each with its timescale small compared to the parameter-drift timescale. The estimation algorithm will use the dataset of each batch to estimate the parameter dynamics. The final output of the current batch, the posterior distribution $p(\mathbf{x}_{0:t}|\mathbf{y}_{1:t})$, will be used as the new prior $p(\mathbf{x}_0)$ of the next batch. Since the timescale of each estimation remains small compared to the parameter-drift timescale, this technique can be considered as a hybrid between online and offline estimation methods.

There are two main advantages over the previous algorithm. First, the partitioned dataset will have smaller trajectory length than one continuous set. This results in a low computational complexity for Eq.~\eqref{eqn:weightUpdateResampling}. Second, it is more efficient than online estimation i.e. estimating the parameter in one continuous take (or as the measurement data arrives). As experimental data are imperfect, the batch estimation allows for averaging signals to increase the effective signal-to-noise ratio. This batch-averaged dataset improves the accuracy of the algorithm while reducing the size of the dataset, making it better than estimating with the raw, noisy dataset.

\section{Map-based Quantum Trajectories for discrete-time data} \label{sec:qTraj}



In this section, we formulate a dynamical model for an open quantum system with continuous measurement, around which we base our SMC algorithm for parameter tracking. A quantum system of interest with its density matrix $\rho$ is assumed to interact with the environment via a finite number of Lindblad operators $\hat c_j$ and a time-continuous observation. In general, time-continuous measurements can be categorized into two types: the jump type (e.g., photon detection) and the diffusive type (e.g., homodyne and heterodyne detections). For this work, we focus on the diffusive type with a single measurement record. However, we note that our formalism can be straightforwardly generalized to accommodate multiple measurement records and the jump process.


Let us consider a diffusive-type homodyne measurement. Its backaction can be described by a diffusive stochastic master equation (SME) conditioned on its measurement result~\cite{BookCarmichael2,BookWiseman, BookJacobs}. Under the strong Markov assumption, the stochastic master equations is given by
\begin{align} \label{eqn:SMEgen}
    \partial_\tau \rho = -i [\hat H, \rho] + \sum_j{\cal D}[\hat c_j]\rho + \dd W {\cal H}[\hat c_{\rm o}]\rho,
\end{align}
where the sum is over the measured and unmeasured channels and $\hat{c}_\mathrm{o}$ is the measured channel. We have omitted the time argument of $\rho(\tau)$ for brevity. The first term in Eq.~\eqref{eqn:SMEgen} describes the system's unitary dynamics, while the second term with ${\cal D}[\hat c]\bullet \equiv \hat c \bullet \hat c^\dagger - \tfrac{1}{2}(\hat c^\dagger \hat c \bullet + \bullet \hat c^\dagger \hat c)$ describes the system's decoherence from environmental coupling via the Lindblad operators $\hat c_j$. The last term describes the measurement backaction via a Lindblad operator $\hat c_{\rm o}$, where ${\cal H}[\hat c]\bullet \equiv \hat c \bullet + \bullet \hat c^\dagger - {\rm Tr}[\hat c \bullet + \bullet \hat c^\dagger]\bullet$ and $\dd W$ is the Wiener increment associated with a measurement result $y(\tau)$ (as a scalar version of the bold-letter vector $\mathbf{y}(\tau)$) through
\begin{align}\label{eqn:signalSME}
y(\tau) {\rm d}\tau = {\rm Tr}[({\hat c}_{\rm o}  + {\hat c}_{\rm o}^\dagger)\rho(\tau)]{\rm d}\tau + {\rm d} W(\tau).
\end{align}
That is, the measurement result is assumed to be a combination of the informative part and the noise.

While the theory assumes the process to be time-continuous, measurement results in the physical world are usually digitized into finite time step denoted by $\ddt$. To reflect reality and be consistent with Section~\ref{sec-smc}, the measurement results will be written as $y_{1:t} \equiv \{y_1, y_2, ... , y_t\}$. The trajectory of the system's density matrices becomes $\rho_{0:t}$, where $\rho_0$ is an initial state. These density matrices contain the information about the system parameters $\mathbf{x}_{1:t}$, where we note that starting with either $\mathbf{x}_0$ or $\mathbf{x}_1$ is simply a matter of bookkeeping.
However, the SME Eq.~\eqref{eqn:SMEgen} can only be applied to a time-continuous record or when the time step is infinitesimally small~\cite{Ralph2011}. We need to use an alternative map-based approach~\cite{Wonglakhon2024} with time resolution $\ddt$ to derive associated operations for unitary dynamics, decoherence, and measurement backaction. The last operation is crucial for computing the observation distribution, $f(y_t|\mathbf{x}_t)$, an important quantity in parameter estimation.

Using the map-based approach, we separate the system's state dynamics into unitary, decoherence, and measurement backaction~\cite{Wonglakhon2024}. We write an update equation for each time step, mapping states from $\rho_{t-1}$ to $\rho_t$, as a concatenated operation (accurate up to $O(\Delta \tau)$): 
\begin{align}\label{eq:evolvestate}
    \rho_{t} \propto {\cal M}(y_t,\mathbf{x}_t)[{\cal Q}(\mathbf{x}_t)[{\cal U}(\mathbf{x}_t)[\rho_{t-1}]]]
\end{align}
where ${\cal U}(\mathbf{x}_t)[\bullet] \equiv e^{-i \hat H(\mathbf{x}_t) \ddt} \bullet e^{+i \hat H(\mathbf{x}_t) \ddt}$ describes the unitary dynamics assuming that the Hamiltonian is constant during the time step and ${\cal Q}(\mathbf{x}_t)[\bullet]$ describes any decoherence irrelevant of the measurement. The last operation ${\cal M}(y_t,\mathbf{x}_t)[\bullet]$ describes the measurement backaction, given a measurement result $y_t$. The reasons for the order of operations is threefolds: Firstly, . Secondly, . It can be written as~\cite{BookWiseman}
\begin{align} \label{eqn:measureMap}
{\cal M}(y_t,\mathbf{x}_t)[\bullet] = \wp_{\rm ost}(y_t){\hat M}(y_t,\mathbf{x}_t) \bullet {\hat M}(y_t,\mathbf{x}_t)^\dagger,
\end{align}
where the \emph{measurement operator} $\hat{M}(y_t, \mathbf{x}_t)$ is
\begin{align}
    \hat{M}(y_t,\mathbf{x}_t)=\hat{1}-\tfrac{1}{2}\op c_{\rm o}^\dagger\op c_{\rm o}\ddt+\op c_{\rm o} y_t \ddt
\end{align}
for the single homodyne measurement associated with the observed Lindblad operator $\hat c_{\rm o}$. The prefactor
$\wp_{\rm ost}(y) \equiv \sqrt{\mathrm{d}\tau/2\pi} \exp(-y^2\mathrm{d}\tau/2)$ is an ostensible probability function~\cite{BookWiseman}, whose purpose is to ensure that an operator given by $\wp_{\rm ost}(y)\hat{M}(y_t,\mathbf{x}_t)^\dagger \hat{M}(y_t,\mathbf{x}_t)$ is a normalized positive-operator-valued-measure element.

To simplify our derivation, we reformulate the concatenated (Markov) operation in Eq.~\eqref{eq:evolvestate} into a Kraus operator $\hat K(y_t,\mathbf{x}_t)$. The update equation becomes
\begin{align}\label{eq-krausmap}
    \rho_{t} = \frac{{\hat K}(y_t,\mathbf{x}_t) \rho_{t-1} {\hat K}(y_t,\mathbf{x}_t)^\dagger}{{\rm Tr}[{\hat K}(y_t,\mathbf{x}_t) \rho_{t-1} {\hat K}(y_t,\mathbf{x}_t)^\dagger]},
\end{align}
where the denominator is the observation distribution $f(y_t|\mathbf{x}_t)$. For the homodyne measurement, the distribution can be approximated as a Gaussian distribution, i.e.,
\begin{align}
  f({y_t} \, |\, & \mathbf{x}_t) =\, {\rm Tr}[\hat{K}(y_t,\mathbf{x}_t)\rho_{t-1}\hat{K}(y_t,\mathbf{x}_t)^\dagger],\label{eq-probygen} \\
  \approx & \frac{1}{\sqrt{2\pi \sigma_t^2}} \exp\left\{-\left[y_t - \mu_t\right]^2/2 \sigma_t^2\right\}. \label{eq-probygen-2}
\end{align}
Their statistical properties (mean and variance) are
\begin{align} \label{eqn:mean}
\mu_t(\mathbf{x}_t)  &= \mathbbm{E}_{y_t|\mathbf{x}_t}[y_t],\\ 
\sigma_t^2(\mathbf{x}_t) &= \mathbbm{E}_{y_t|\mathbf{x}_t}[y_t^2] - (\mathbbm{E}_{y_t|\mathbf{x}_t}[y_t])^2. \label{eqn:var}
\end{align}
where $\mathbbm{E}_{y_t|\mathbf{x}_t}$ refers to expectation value computed from the observation distribution $f(y_t|\mathbf{x}_t)$. It is useful to note that the mean value can be written in terms of the concatenated operation: $\mu_t = \int f(y_t|\mathbf{x}_t)y_t \mathrm{d} y_t \approx {\rm Tr}[(\hat c_{\rm o} + \hat c_{\rm o}^\dagger) {\cal Q}(\mathbf{x}_t)[{\cal U}(\mathbf{x}_t)[\rho_{t-1}]]]$ and the variance is simply $\sigma_t^2 = 1/\ddt$, to the lowest order in $\ddt$.

If a single run of experiment with measurement results $y_{1:t}$ contains enough information about the unknown parameters $\mathbf{x}_{1:t}$, the observation distribution in Eq.~\eqref{eq-probygen-2} is adequate for the SMC algorithms. However, most experiments have highly noisy measurement results. One can consider performing multiple runs of the same experiment and averaging multiple measurement results to increase the signal-to-noise ratio, assuming that the unknown parameters do not change appreciably over that timescale. Let us define the average measurement result
\begin{align}\label{eq-averecord}
    {\bar y}_t \equiv \frac{1}{\ns} \sum_{j=1}^{\ns} y_{t,j},
\end{align}
where $\ns$ is the number of measurement results. We note that $\ns$ can vary for different time steps $t$. We assume constant $\ns$ for the ease of bookkeeping. In order to use the average measurement result $\bar{y}_t$ in place of $y_t$ for the SMC algorithm, we introduce a new map $\rho_t = {\cal K}(\bar y_t,\mathbf{x}_t)[\rho_{t-1}]$ similar to Eq.~\eqref{eq-krausmap} such that
\begin{align}\label{eqn:avgMap}
    {\cal K}(\bar y_t,\mathbf{x}_t)[\rho] =\!\! \frac{1}{\ns} \sum_{j=1}^{\ns} \frac{\hat K(y_{t,j},\mathbf{x}_t) \rho \hat K( y_{t,j},\mathbf{x}_t)^\dagger}{{\rm Tr}[\hat K(y_{t,j},\mathbf{x}_t) \rho \hat K(y_{t,j},\mathbf{x}_t)^\dagger]}
\end{align}
is true and applied to any density matrix $\rho$, given that $\mathbf{x}_t$ does not vary with $j$. We note that this is an interpolation between a single-trajectory Kraus map and the Lindblad evolution, which can be obtained from summing all possible trajectories. Appendix~\ref{app:fullMap} shows an approximated form of Eq.~\eqref{eqn:avgMap} up to $O(\ddt^2)$. Using Gaussian properties, it is easy to convince ourselves that the observation distribution for the average signal is
\begin{align} \label{eqn:realObsDist}
  f({\bar y}_t \, |\, \mathbf{x}_t) \approx  \frac{1}{\sqrt{2\pi {\bar\sigma}_t^2}} \exp\left\{-\left[{\bar y}_t - \mu_t\right]^2/2 {\bar\sigma}^2\right\},
\end{align}
where the mean $\mu_t$ stays the same as in Eq.~\eqref{eqn:mean}, but the variance is scaled to $\bar \sigma^2 = \sigma_t^2/\ns= (\ns\ddt)^{-1}$.

For our work, we will use the average signals in Eq.~\eqref{eq-averecord} for the SMC algorithm and the observation distribution in Eq.~\eqref{eqn:realObsDist}. Therefore, the weight update Eq.~\eqref{eqn:ourWeightUpdate} for each particle $(i)$ becomes
\begin{align} \label{eqn:qTWeight}
{\tilde w}_t^{(i)} \propto \exp \left[-\left ( {\bar y}_t - \mu_t^{(i)} \right)^2/(2{\bar\sigma}^2) \right ] {\tilde w}_{t-1}^{(i)}.
\end{align}
The update equation Eq.~\eqref{eqn:weightUpdateResampling} in the resampling process also takes a similar form.

\section{Implementing sequential Monte Carlo algorithms}\label{sec-implement}

With the SMC algorithms introduced in Section~\ref{sec-smc} and the quantum trajectories in Section~\ref{sec:qTraj}, we are now ready to implement the estimation algorithms on the experimental data. We consider data from superconducting qubits coupled to microwave cavities because of their impressively fast detection rate compared to their parameter-drift timescale.
We choose datasets from two different superconducting-qubit setups: the dispersive $z$-measurement setup from Weber et al. (2014)~\cite{Weber2014} and the fluorescence-measurement setup from Naghiloo et al.~\cite{Naghiloo2016, Naghiloo2017}. The datasets are homodyne-type measurement signals obtained from near-quantum-limited Josephson parametric amplifiers with multiple runs of $T$-duration trajectories. Since the multiple runs of experiments occurred one after another for a period of time, we would expect the experimental parameters to possibly drift over that period. We note that, in the latter setup, there exists multiple datasets with different parameters, which we can use to test how the SMC algorithms may estimate parameter values from the dataset alone.

In the following subsections, we explain our algorithm step-by-step, then estimation results from the two superconducting-qubit datasets. We start with the fluorescence setup, as it contains many more data points, and then follow with the $z$-measurement setup, which contains fewer data points. The results from both datasets are compared with independently calibrated parameters. We show that our algorithm has comparable accuracy to the conventional calibration and that it can detect hidden dynamics inaccessible to the conventional method.

\subsection{Step-by-step algorithms}\label{subsec:algoDetail}

In this section, we summarize the algorithms presented in Sections~\ref{subsec:combinedSMC}~and~\ref{subsec:longSMCS}. We begin by offsetting and rescaling the raw measurement signals $V_t$ to the appropriate units according to the formalism of Section~\ref{sec:qTraj}. That is, the rescaled signals $v_t$,
\begin{align} \label{eqn:resEqn}
    v_t \propto (V_t - V_{\rm off}),
\end{align}
where $V_\mathrm{off}$ is the chosen offset value.
We then partition the dataset into multiple batches, each with $\ns$ trajectories. By averaging trajectories of each batch, we obtain a batch-averaged signal: $\bar{v}_t = \sum_j^{\ns}v_{t,j}/\ns$. To account for errors in the offset value in Eq.~\eqref{eqn:resEqn}, an additional unknown parameter $v_{\mathrm{off}, t}$ is introduced to the measurement result, i.e., we define the batch-averaged measurement result as
\begin{align} \label{eqn:result}
    \bar{y}_t = \bar{v}_t - v_{\mathrm{off}, t}
\end{align}
to be used in place of Eq.~\eqref{eq-averecord}. We note that, from now on, we will use the superscript `$(i)$' to index the SMC particles. Thus, the measurement result for each particle is $\bar{y}_t^{(i)} = \bar{v}_t - v_{\mathrm{off}, t}^{(i)}$. This completes our data pre-processing.

Given a series of batch-averaged signal, $\bar{v}_{0:t}$, we implement the SMC algorithm to estimate the unknown parameters $\mathbf{x}_{0:t}$ of interest (including the unknown offset $v_{\mathrm{off}, 0:t}$). The algorithm runs as follows:
\begin{enumerate}
    \item \emph{Initialization}: At the first time step $t=0$, $\np$ particles of initial parameter values $\mathbf{x}_0^{(i)}$ and density matrices $\rho_0^{(i)}$ from pre-defined prior distributions are initialized. All particles are assigned with equal weights $\tilde{w}_0^{(i)} = 1/\np$.
    \item \emph{Evolution}: For any step $t$, measurement results $\bar{y}_t^{(i)}$ for each particle $i$ are calculated from Eq.~\eqref{eqn:result} and associated density matrices are evolved using Eq.~\eqref{eqn:avgMap}. Then, the mean signal $\mu_t^{(i)}$ and variance $(\bar{\sigma}^{(i)}_t)^2$ are obtained using Eq.~\eqref{eqn:mean} and Eq.~\eqref{eqn:var}, respectively.
    \item \emph{Weight Update}: The weight of each particle is calculated using Eq.~\eqref{eqn:qTWeight}, which can be normalized to obtain $\tilde{w}_t^{(i)} = w_t^{(i)}/\sum_j w_t^{(j)}$. Then,  the empirical mean of a desired function $h(\mathbf{x}_{0:t})$ is obtained with Eq.~\eqref{eqn:empMean}. In our case, $h(\mathbf{x}_{0:t}) = \mathbf{x}_{0:t}$, its empirical mean will be the estimator of unknown parameters of interest. Its error can be obtained using the weighted standard deviation. 
    \item \emph{Resampling}: The weights from the previous step are used to compute $\neff=1/\sum_i (\tilde{w}_t^{(i)})^2$. If $\neff < \nth$, the resampling step is initialized, as described in Section~\ref{subsec:combinedSMC}.
    \item Step 2 through 4 are repeated until the final time step $t=n$ is reached. Once the signals in a given batch are processed, the final distribution of particles and their weights will be used as the new empirical prior distribution for the next batch. These steps are repeated until all signal data are processed.
\end{enumerate}

We note here that the algorithm requires specifying \emph{hyperparameters}, defined to be the number of signal trajectories per batch $\ns$, the number of particles $\np$, the threshold for resampling $\nth$, and the resampling ratio $p,q$. Their values can be deliberately chosen to achieve high efficiency. For our work, we determine their values by testing different sets of hyperparameter values with numerical simulations. We simulate the physical systems and their measurement processes to approximate the algorithm's effectiveness and compare the impact of the hyperparameter values. For the detail of the simulation for the superconducting-qubit datasets, see Appendix~\ref{app:simEst}.


Finally, to assess the accuracy of our estimation, we introduce a method that can be applied even when true values of the parameters are inaccessible. We call this method the \emph{signal reconstruction} method. For each batch, this can be done by comparing sample-averaged raw measurement signals, $\bar{V}_t$, obtained directly from experiments, and averaged signals that are numerically simulated from the system's dynamical model using the estimated parameters. 
In obtaining the averaged signal $\check{V}_t$, we first simulate averaged measurement result $\bar{y}_t$. This can be done in two ways: using the Kraus map, generating $\ns$ stochastic measurement results $y_{t,j}$ and their associated states via Eq.~\eqref{eqn:signalSME} and Eq.~\eqref{eq:evolvestate}, where the average result $\bar{y}_t$ is obtained from Eq.~\eqref{eq-averecord}, or using the Lindblad master equation and its averaged measurement result, given by Eq.~\eqref{eqn:SMEgen} and  Eq.~\eqref{eqn:signalSME}, but without the stochastic ${\rm d}W$ terms.
By plotting the averaged signals from the experiment and the simulations, we may see their differences and similarities. Moreover, we can quantify the difference more rigorously by computing the root-mean-square error,
\begin{equation}\label{eq:rmse}
    \mathrm{RMSE} = \left [\frac{1}{n} \sum_t^n (\bar{V}_t - \check{V}_t)^2 \right ]^{1/2},
\end{equation}
where $\bar{V}_t$ is the sample-averaged raw signal and $\check{V}_t$ is the simulated (averaged) signal. However, in the following sections, we will only compute the RMSE from the Lindblad master equation, simply to avoid any unnecessary errors from stochastic simulation.

\subsection{Experimental data: Fluorescence measurement of transmon qubit} \label{subsec:fluo}

\begin{figure}
    \centering
    \includegraphics[width=\linewidth]{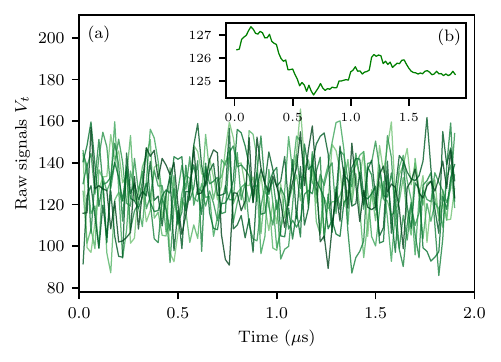}
    \caption{Examples of raw signals $V_t$ and their sample mean. (a) shows randomly chosen 8 full-length raw signals $V_t$. (b) shows the sample-mean signal $\bar{V}_t$ averaged over $10\,000$ full-length signals. The sample mean reveals some hidden dynamics, from which some information about the system's parameters can be extracted.}
    \label{fig:rawSignal}
\end{figure}

\begin{figure*}[t]
  \includegraphics[width=\textwidth]{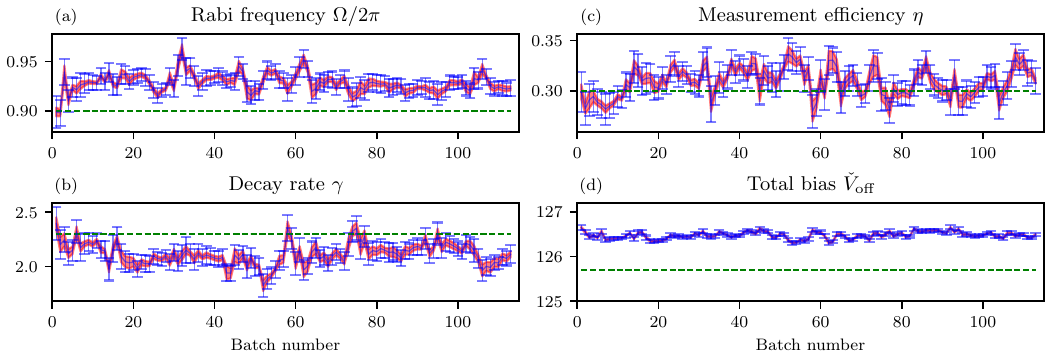}
  \caption{Estimation results of four unknown parameters from the transmon qubit under  the fluorescence measurement setting: (a) Rabi frequency $\Omega/2\pi$, (b) fluorescence decay rate $\gamma$, (c) measurement efficiency $\eta$, and (d) total signal bias $\check V_{\rm off}$. 
  For each parameter, the estimated values are shown with their uncertainties for all 113 batch numbers. The measurement duration runs for approximately $1.88 \times 113 \times \ns \approx 2.2 \, {\rm second}$. Each data point is obtained by averaging the estimates over $24$ runs. The red shaded region reveals the standard deviation of the mean. The blue error bars show the parameter uncertainties, which are calculated from averaging the weighted standard deviation of each of the $24$ runs. The green dash lines are calibrated values obtained from Naghiloo et al. (2016)~\cite{Naghiloo2016} and Naghiloo et al. (2017)~\cite{Naghiloo2017}.}
  \label{fig:allParam_fluo}
\end{figure*}

We start with the superconducting-qubit experiment in Refs.~\cite{Naghiloo2016, Naghiloo2017}. The setup was a superconducting transmon qubit coupled to a microwave cavity, where the qubit was coherently driven with a Rabi drive $\Omega$. The qubit-cavity coupling was tuned such that the qubit experienced relaxation decay with a rate $\gamma$. The decay was monitored via parametric amplifier, which constituted a homodyne measurement with an efficiency $\eta$.

Formally, its dynamics is effectively governed by a stochastic master equation similar to Eq.~\eqref{eqn:SMEgen},
\begin{align} \label{eqn:SMEfluo}
    \partial_t\rho = -i [\tfrac{\Omega}{2}\hat\sigma_y, \rho] + \gamma {\cal D}[\hat \sigma_-]\rho  + \sqrt{\gamma \eta} \, {\rm d}W{\cal H}[\hat \sigma_-]\rho,
\end{align}
where $\hat{\sigma}_{x}$ and $\hat{\sigma}_{y}$ are the $x$ and $y$ Pauli matrices respectively and $\hat{\sigma}_- = (\hat{\sigma}_x -i\hat{\sigma}_y)/2$ is the qubit's decay operator. The Wiener increment ${\rm d}W(\tau)$ is related to the measurement result via 
\begin{align}
    y(\tau) = \sqrt{\eta\gamma}\,{\rm Tr}[\hat{\sigma}_x \rho(\tau)] + {\rm d}W(\tau).
\end{align}
This corresponds with the observed Lindblad operator $\hat c_{\rm o} = \sqrt{\eta\gamma}\, \hat\sigma_-$ in Eq.~\eqref{eqn:signalSME}.

The experimental dataset from Refs.~\cite{Naghiloo2016, Naghiloo2017} contains $1\,159\,893$ trajectories of raw (voltage measurement) signals with varying time lengths, ranging from $1$ step to at most $94$ steps with the time step $\ddt = 0.02$ $\mu$s. Thus, the longest trajectory time is $T = 1.88$ $\mu$s. The raw signals can be rescaled into the rescaled signals $v_t$ using
\begin{equation}
    v_t = \frac{V_t - V_\mathrm{off}}{\Delta V \ddt},
\end{equation}
where $\Delta V = 137.3$ and $V_\mathrm{off} = 125.4$ are the scaling factor and the baseline offset from the detector calibration. Adding the residual bias $v_\mathrm{off}$ via Eq.~\eqref{eqn:result}, the total bias becomes $\check{V}_\mathrm{off} = V_\mathrm{off} + v_\mathrm{off}\Delta V\ddt$. 
We note that our unit is different from the ones in Ref.~\cite{Naghiloo2016, Naghiloo2017}. Lastly, independent calibration shows the initial state of the transmon qubit to be at Bloch coordinate $\mathbf{r}_0 = (-0.07,0,0.96)$~\cite{Naghiloo2017}.

In Section~\ref{subsec:longSMCS}, we discussed how the batch estimation is useful for the case of a large and noisy dataset. Here, we show examples of raw signals $V_t$ and their sample mean in Figure~\ref{fig:rawSignal}. After averaging the noisy individual trajectories, the sample mean reveals some hidden dynamics, from which the algorithm may extract information about the system parameters. Thus, it is convincing that we can obtain a better signal-to-noise ratio by averaging multiple trajectories into one. 

As the SMC algorithm works in the discrete regime, the continuous dynamics of Eq.~\eqref{eqn:SMEfluo} must be modified into a discrete operation in accordance to Section~\ref{sec:qTraj}. According to the formalism,
\begin{subequations} \label{eqn:fluoMap}
\begin{align}
\mathbf{x}_t =&\, (\Omega_t, \gamma_t, \eta_t, v_{\mathrm{off}, t}) ,\label{eqn:fluoMap1}\\
    \hat H(\mathbf{x}_t) =&\, \frac{\Omega_t}{2}\hat\sigma_y ,\\
    {\cal Q}(\mathbf{x}_t)[\rho] =& \,(1 - \eta_t)\gamma_t\ddt{\cal D}[\hat \sigma_-]\rho  ,\\
    {\hat M}(y_t, \mathbf{x}_t) =&\, \hat 1 - \frac{\eta_t\gamma_t}{2}\hat \sigma_+ \hat \sigma_- \ddt + y_t \sqrt{\eta_t \gamma_t} \, \hat \sigma_- \ddt,
\end{align}
\end{subequations}
noting that $\hat \sigma_+ = \hat\sigma_x + i\hat \sigma_y$. The Kraus superoperator $\mathcal{K}(\bar{y}_t, \mathbf{x}_t)$ can be constructed via Eq.~\eqref{eq-krausmap} from the measurement operator $\hat{M}(y_t,\mathbf{x}_t)$. For an efficient approximated map, see Eq.~\eqref{eq-expandmap}. Every parameter has the subscript $t$ to leave open the probability that their value may change over time, although we keep the assumption that the system has slow-varying dynamics.

With the dynamical model prepared, we now choose the appropriate hyperparameter values for our algorithm.
This is done using numerical simulations, whose comparison results are discussed in Appendix~\ref{app:simEst}. In summary, we take the number of trajectories per batch to be $\ns = 51 \times 21 = 10\,200$, the number of particles $\np = 1024$, the threshold $\nth=N_p/2$, the ratio for defensive strategy $p=0.9$, and the ratio for Gaussian variance $q=0.1$. The chosen $p$ and $q$ indicates that approximately $0.9\np$ particles will have the kernel's variance of $0.1\Sigma_t$, while the remaining $0.1\np$ particles will have the variance $\Sigma_t$. We note that the reason for the specific value of $\ns$ stems from the specific structure of our dataset; its signal trajectories have varying lengths, in blocks of $51$ trajectories. From Appendix~\ref{app:simEst}, a reasonable number of trajectories per batch is $10\,000$, leading to choosing $\ns=10\,200$ since it is a round multiple of $51$. 

We test the algorithm's stability and explore statistical details of the estimated parameters by running the algorithm repeatedly $24$ times. Figure~\ref{fig:allParam_fluo} shows the estimation results, plotted over $1\,159\,893/\ns \approx 113$ batches. The parameters of interest are those in Eq.~\eqref{eqn:fluoMap1}. Each data point represents the estimated parameters averaged over the $24$ runs, with the red shaded region showing the standard deviation of the mean. The blue error bars show the parameter uncertainties, which are calculated from averaging the weighted standard deviation of each of the $24$ runs. The fact that the red shaded regions always fit inside the blue error bars indicates that our SMC algorithm gives reliable estimators, within the uncertainties predicted by the weights. 
Next, we compare our results with experimentally calibrated values, plotted as green dash lines. Due to incomplete information, we have used values from two independent calibrations. The first calibration is from Naghiloo et al. (2016)~\cite{Naghiloo2016}, where we have used the decay rate $\gamma$, the measurement efficiency $\eta$, and the total bias $\check{V}_\mathrm{off}$. The other calibration is from Naghiloo et al. (2017)~\cite{Naghiloo2017}, where we have used Rabi frequency $\Omega$. The second calibration is on a different transmon qubit and has different values from our estimation, but it is the only test that has the estimate for $\Omega$. From Figure~\ref{fig:allParam_fluo}, our estimates are overall consistent with the range of calibrated values, but they are not in perfect agreement. Our $\Omega$ estimates (a) are slightly above the calibrated value (green dash) most of the time, while, our $\gamma$ estimates (b) are slightly below its calibrated value. The estimates for $\eta$ (c) agree quite well with the calibrated value. Interestingly, for the total bias $\check{V}_\mathrm{off}$ (d), even though its estimated value and its calibrated value fall within a similar range of 125 to 127 (still not a significant difference in experiments), the discrepancy between the two is much larger than the uncertainties of the estimates. This indicates the SMC algorithm's confidence in its estimation, raising questions about the reliability of the calibrated value. 

Table~\ref{tab:estCalComp_fluo} compares the estimates of the final ($113$th) batch to the calibrated values from two different references. The estimates agree partially with the calibration, with $\eta$ from Ref.~\cite{Naghiloo2016} and $\Omega/2\pi$ from Ref.~\cite{Naghiloo2017}. However, the estimates for $\gamma$ and $\check{V}_\mathrm{off}$ are only moderately in range with Ref.~\cite{Naghiloo2016}. We note that the calibrated values may not represent the true values. The calibration process relies on strong measurement and regression for individual parameter, making it susceptible to falling into local minima that may not necessarily be the global minimum. More importantly, strong measurements cannot be done concurrently with the signal-acquisition process. The values obtained may not reflect the parameter values at the time when the data are gathered.

\begin{table}[t]
\begin{tabular}{crrr}
\hline
\multicolumn{1}{l}{} & \multicolumn{1}{c}{Estimation} & \multicolumn{1}{c}{Ref. \cite{Naghiloo2016}} & \multicolumn{1}{l}{Ref. \cite{Naghiloo2017}} \\ \hline
$\Omega/2\pi$ & $0.92\pm0.01$ & N/A & $0.9$ \\
$\gamma$ & $2.11\pm0.08$ & $2.3$ & $1.42$ \\
$\eta$ & $0.31\pm0.01$ & $0.3$ & $0.45$ \\
$\check{V}_\mathrm{off}$ & $126.2 \pm 0.1$ & $125.694$* & N/A \\ \hline
\end{tabular}
\caption{Comparison between the final estimated values of the parameters and the calibrated values. *This value of $\check{V}_\mathrm{off}$ is obtained from direct communication with the authors of Refs.~\cite{Naghiloo2016,Naghiloo2017}.}
\label{tab:estCalComp_fluo}
\end{table}

\begin{figure}
    \centering
    \includegraphics[width=\linewidth]{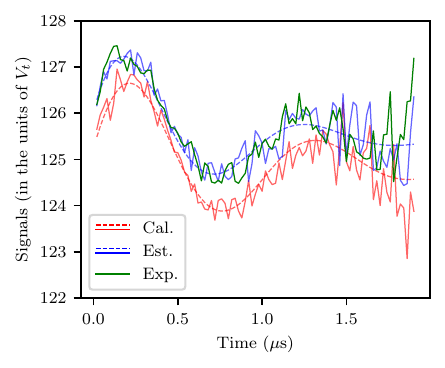}
    \caption{Signal reconstruction for the fluorescence measurement dataset. We use the signals and estimated parameters from the final (113th) batch for the reconstruction. The green curve is the batch-averaged raw signal. The red and blue colored curves are simulated (averaged) signals using parameters from the calibration (Cal.) and from our SMC estimation (Est.) respectively. The dash and solid features indicate that the simulated signals are obtained from the Lindblad master equation and from the Kraus maps respectively. }
    \label{fig:sigRecon_fluo}
\end{figure}

Lastly, we use the signal reconstruction method introduced in Section~\ref{subsec:algoDetail}. Figure~\ref{fig:sigRecon_fluo} compares the averaged raw signal (green), the simulated signals from SMC-estimated parameters (blue), and ones from calibrated parameters (red). For the latter two, there are solid and dash lines, showing the simulated signals from the stochastic Kraus-map simulation (jacked lines) and from the Lindblad master equation's solutions (smooth curves) respectively. We have used the calibrated values from Ref.~\cite{Naghiloo2016}, except $\Omega$, which we obtain from Ref.~\cite{Naghiloo2017}.  It is easy to see that the signals reconstructed from the SMC-estimated parameters are closer to the raw signals than those from the calibration. We calculate RMSE with Eq.~\eqref{eq:rmse} for both the SMC algorithm and the calibration respectively: $\mathrm{RMSE}_\mathrm{est} = 0.3699$ and $\mathrm{RMSE}_\mathrm{cal} = 0.7465$, confirming that our algorithm produces better estimates for the parameters than the standard calibration process.

\subsection{Experimental data: Dispersive $z$ measurement of transmon qubit} \label{subsec:zMeas}

\begin{figure*}[t]
    \centering
    \includegraphics[width=\linewidth]{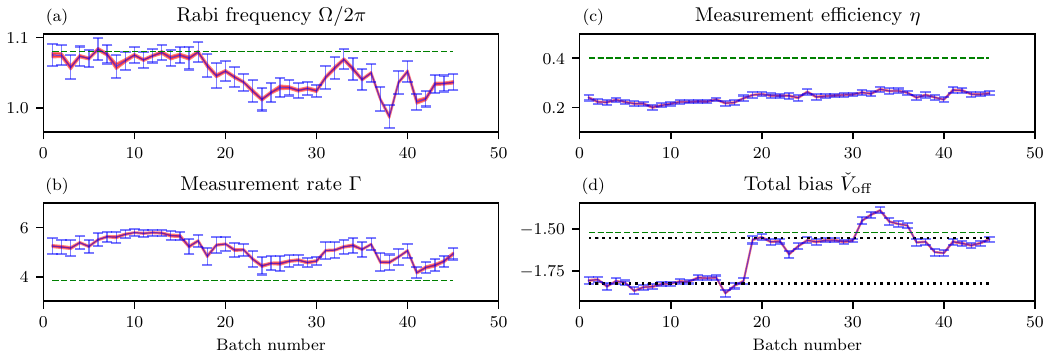}
    \caption{Estimation results of four unknown parameters from the transmon qubit under the dispersive $z$-measurement setting: (a) Rabi frequency $\Omega/2\pi$, (b) measurement rate $\Gamma$, (c) measurement efficiency $\eta$, and (d) total signal bias $\check V_{\rm off}$. The estimated parameters are shown with their uncertainties for all $45$ batch numbers. The measurement duration is approximately $1.424 \times 45 \times \ns \approx 0.3 \, {\rm second}$. Each data point is obtained by averaging the estimates over $24$ runs. The red shaded region and the blue error bars are uncertainties as defined in Figure~\ref{fig:allParam_fluo}. The green dash lines are calibrated values from Weber et al.~\cite{Weber2014}.
    For total bias $\check{V}_\mathrm{off}$, we have split the estimation into two time periods: the first 17 batches and the rest. The black dotted lines show time-averaged estimates of the two periods separately.}
    \label{fig:allParam_zMeas}
\end{figure*}

For the second dataset, we work with the superconducting-qubit experiment in Weber et al.~\cite{Weber2014}. The setup was a superconducting transmon qubit coupled to a microwave cavity, but in a dispersive regime. The qubit was coherently driven with a Rabi drive $\Omega$, while the dispersive cavity was monitored via parametric amplifier, resulting in a continuous measurement of the qubit's $\hat \sigma_z$ with a measurement rate $\Gamma$ and an efficiency $\eta$. Formally, its dynamics is governed by a stochastic master equation,
\begin{equation} \label{eqn:SME_z}
    \partial_t\rho = -i[\tfrac{\Omega}{2}\hat{\sigma}_y, \rho] + \tfrac{\Gamma}{2}\mathcal{D}[\hat{\sigma}_z]\rho + \sqrt{\tfrac{\eta\Gamma}{2}} {\rm d}W \mathcal{H}[\hat{\sigma}_z]\rho.
\end{equation}
The Wiener increment ${\rm d}W(\tau)$ is related to the measurement result via
\begin{align}
    y(\tau) = \sqrt{2\eta \Gamma}\, {\rm Tr}[\hat \sigma_z \rho(\tau)] + {\rm d}W(\tau).
\end{align}
This corresponds with the observed Lindblad operator $\hat c_{\rm o} = \sqrt{\eta\Gamma/2}\, \hat\sigma_z$ in Eq.~\eqref{eqn:signalSME}. We note that we have neglected the effect of the relaxation time $T_1$, whose contribution could result in another term, $(1/T_1)\mathcal{D}[\hat{\sigma}_-]\rho$, in Eq.~\eqref{eqn:SME_z}, as the trajectory duration was too short for the process to have appreciable effect.

The experimental dataset from Ref.~\cite{Weber2014} contains $181\,192$ trajectories of raw (voltage measurement) signals. Unlike those from the previous example, these signal trajectories all have $90$ time steps with $\ddt = 0.016$ $\mu$s. Therefore, the trajectory time is $T = 1.424$ $\mu$s. The raw signals can be rescaled into the rescaled signals $v_t$ using
\begin{equation}
    v_t = \frac{V_t - V_\mathrm{off}}{\Delta V}
\end{equation}
where $V_\mathrm{off} = -1.52$ and $\Delta V = 2.055$. Accounting for the residual bias $v_\mathrm{off}$, we have the total bias $\check{V}_\mathrm{off} = V_\mathrm{off} + v_\mathrm{off}\Delta V$. Finally, independent calibration shows the initial state of the transmon qubit to be at $\mathbf{r}_0 = (0.88,0,0)$~\cite{Weber2014}.

As with the previous section, the dynamics of the system is modified into a Kraus operation. Using the language of Section~\ref{sec:qTraj}, we obtain
\begin{subequations}\label{eqn:diszMap}
\begin{align}
    \mathbf{x}_t &= \{\Omega_t, \Gamma_t, \eta_t, v_{\mathrm{off}, t} \}, \\
    \hat{H}(\mathbf{x}_t) &= \frac{\Omega_t}{2}\hat{\sigma}_y, \\
    \mathcal{Q}(\mathbf{x}_t)[\rho_t] &= \rho_t + (1-\eta_t)\frac{\Gamma_t}{2} \ddt \mathcal{D}[\hat\sigma_z]\rho_t, \\
    \hat{M}(y_t, \mathbf{x}_t) &= \hat{1} - \frac{\eta_t \Gamma_t}{4}\hat 1\ddt + y_t\sqrt{\frac{\eta_t\Gamma_t}{2}}\hat\sigma_z\ddt.
\end{align}
\end{subequations}
Similar to the previous subsection, the Kraus superoperator ${\cal K}(\bar{y}_t, \mathbf{x}_t)$ can be constructed via Eq.~\eqref{eq-krausmap}, which is approximated as in Eq.~\eqref{eq-expandmap}.

The choices of hyperparameters for this dataset are analyzed in Appendix~\ref{app:simEst}. As with the previous case, this is done using numerical simulations. In summary, we choose $\ns = 4000$, $\np = 2048$, $\nth = \np/2$. As for the scale parameter of the Gaussian kernel in the resampling step, we choose $p=0.9$ and $q=0.1$, i.e., approximately $0.9\np$ particles will have the kernel's variance of $0.1\Sigma_t$, while the rest will have the variance of $\Sigma_t$.

Like the previous example, we run our SMC algorithm $24$ times to test its stability and explore statistical details of the estimated parameters. Figure~\ref{fig:allParam_zMeas} shows the estimation results for all $181\,192/\ns \sim 45$ batches. The data points, the red shaded regions, and the blue error bars are the mean of estimators, the standard deviation of the mean, and weighted uncertainties, as defined in the previous subsection. The results show two interesting features. First, the red shaded regions and the blue error bars are both small, indicating the algorithm's strong confidence in the estimated parameters. Second, there exist parameter-drift features, particularly clear in the Rabi frequency $\Omega/2\pi$ (a) and the signal total bias $\check{V}_\mathrm{off}$ (d). One can see that for the first $17$ batches, all estimated parameters are quite stable, even if they do not agree perfectly with the calibration values (green dash lines). However, after (roughly) the 17th batch, the estimates of the Rabi frequency, the decay rate, and especially the signal total bias show more fluctuation over time. The total bias (d) shows a clear jump in values. To investigate $\check{V}_\mathrm{off}$ further, we divide its estimated values in two parts, before and after the $17$th batch, and averages them, obtaining $-1.82 \pm 0.02$ and $-1.55 \pm 0.02$ respectively. The averaged values are shown as the two black dotted lines in Figure~\hyperlink{fig:allParam_zMeas}{5.d}. It is interesting to note that the average value after the $18$th batch is close to the calibration value of $-1.52$~\cite{Weber2014}. This indicates that the calibration could fail to capture any time-dependent dynamics of estimated parameters, which can now be resolved with our algorithm. 

Table~\ref{tab:paramCompzMeas} compares the final ($45$th batch) estimates to the calibrated values from Weber et al.~(2014)~\cite{Weber2014}. The Rabi frequency $\Omega/2\pi$ and the total bias $\check{V}_\mathrm{off}$ show good agreement, while the values for $\Gamma$ and $\eta$ are still quite different, which will be investigated further below. 

\begin{table}[t]
\begin{tabular}{@{}crr@{}}
\toprule
\multicolumn{1}{l}{} & \multicolumn{1}{c}{Estimation} & \multicolumn{1}{c}{Ref.~\cite{Weber2014}} \\ \midrule
$\Omega/2\pi$ & $1.04 \pm 0.01$ & $1.08$ \\
$\Gamma$ & $4.93 \pm 0.23$ & $3.85$ \\
$\eta$ & $0.26 \pm 0.01$ & $0.41$ \\
$\check{V}_\mathrm{off}$ & $-1.56 \pm 0.01$ & $-1.52$ \\ \bottomrule
\end{tabular}
\caption{Comparison between the final estimated values for the parameters and the calibrated values. The calibrated values are obtained from Ref.~\cite{Weber2014}.}
\label{tab:paramCompzMeas}
\end{table}

\begin{figure}[t]
    \centering
    \includegraphics[width=\linewidth]{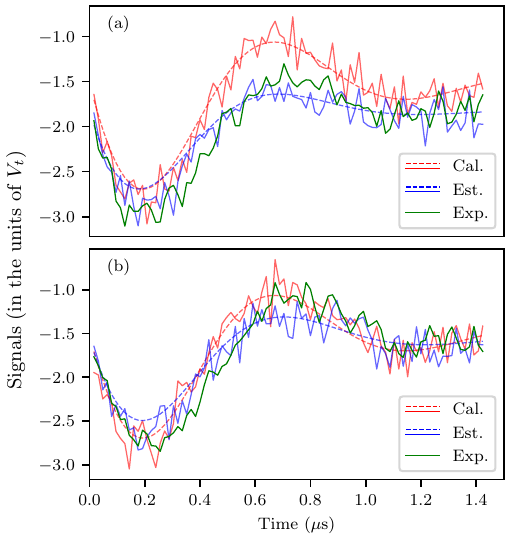}
    \caption{Signal reconstruction for the $z$-measurement dataset. We use the signals and estimated parameters from the 17th batch in (a) and the final (45th) batch in (b) for the reconstruction. The color coding is the same as in Figure~\ref{fig:sigRecon_fluo}. The green curve is the batch-averaged raw signal. The red or blue colored curves are simulated (averaged) signals using parameters from the calibration (Cal.) or from our SMC estimation (Est.) algorithm respectively.
    }
    \label{fig:sigRecon_z}
\end{figure}

\begin{figure}[t]
    \centering
    \includegraphics[width=\linewidth]{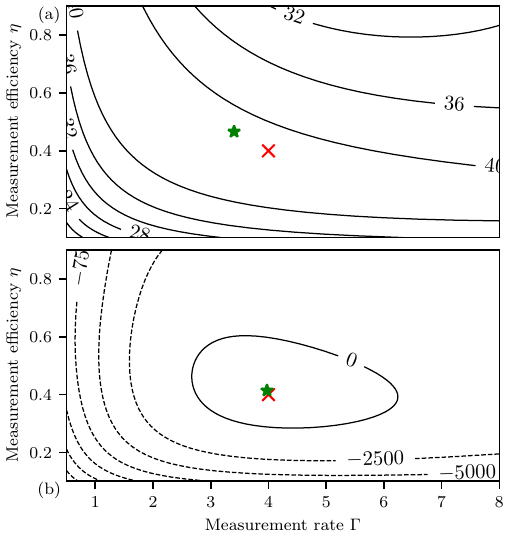}
    \caption{The log-likelihood contour surfaces of the parameter pair $(\eta, \Gamma)$ for the $z$-measurement setting, where the red marks show the true parameter values and the green stars are estimators. Top, (a), is for low numbers of time steps and signals per batch ($n = 10, N_s = 1000$); Bottom, (b), for high numbers of time steps and signals per batch ($n = 90, N_s = 200\,000$). These graphs are used in investigating the parameter identification problem in estimating $\Gamma$ and $\eta$.}
    \label{fig:contour_z}
\end{figure}

For our signal reconstruction test, we divide the test into two parts: using the estimates from the $17$th batch and the final (45th) batch, shown in Figure~\hyperlink{fig:sigRecon_z}{5.a} and \hyperlink{fig:sigRecon_z}{5.b} respectively. The figure uses same plotting styles and legends as in Figure~\ref{fig:sigRecon_fluo}.
It is clear that, in (a), the estimated values can describe the truth (raw signals) better than the calibrated values. 
However, in (b), both the estimated and calibrated values produce signals with comparable quality. We calculate RMSE to quantify their accuracy. For part (a), we get $\mathrm{RMSE}_{\mathrm{est}, (a)} = 0.1958$ and $\mathrm{RMSE}_{\mathrm{cal}, (a)} = 0.3540$. For part (b), we get $\mathrm{RMSE}_{\mathrm{est}, (b)} = 0.2085$ and $\mathrm{RMSE}_{\mathrm{cal}, (b)} = 0.1984$, confirming that our SMC algorithm does better than the calibration method for part (a) and does just as well as the calibration for (b).


Let us investigate the estimation of $\Gamma$ and $\eta$ further. From Appendix~\ref{app:simEst}, we find that the estimated values of $\Gamma$ and $\eta$ do not converge to the true values. This can be attributed to the parameter identification problem. For certain models, multiple parameter values can generate similar sets of measurement results. An inspection of Eqs.~\eqref{eqn:SME_z} and \eqref{eqn:diszMap} reveals that the two parameters almost always appear as a product $\beta = \eta\Gamma$, meaning that the algorithm likely estimates the product instead of the individual parameters. If one attempts to estimate the parameters individually, random drift in one parameter may cause the other to change to compensate.
As a demonstration, we calculate the log-likelihood contour surface of the two parameters, as shown in Figure~\ref{fig:contour_z}. The contour surface is calculated by varying the values of $(\eta, \Gamma)$ while fixing the rest. 
The red markers show the true values of parameters, while the green stars are the estimated values from averaging across the contour surfaces. Figure~\hyperlink{fig:contour_z}{6.a} shows the log-likelihood contours, $\log \prod_{i=1}^{t} f(\bar{y}_i|\Gamma, \eta)$ for a batch-averaged measurement signal with only $t=10$ time steps and $\ns = 1000$ per batch. It clearly shows that the contour lines are approximately hyperbolic, consistent with the fact that the algorithm estimates the product $\beta = \eta\Gamma$. On the other hand, Figure~\hyperlink{fig:contour_z}{6.b} is for a batch-averaged signal with $t=90$ time steps and $\ns=200\,000$ per batch. The latter estimation is much better than the former case, likely because of two reasons. First is that the large average size per batch $\ns$ should reduce noise and refine the estimation. Second, from a careful inspection that there is still one term in Eq.~\eqref{eqn:SME_z} which $\Gamma$ appears alone, the longer time steps should accumulate more information about the individual parameters.

The parameter identification problem only explains why the algorithm may not converge to correct values. It does not explain the parameter drift that we have seen in Figure~\ref{fig:allParam_zMeas}, for $\Gamma$ and $\eta$. 
There are many possible reasons, including the unaccounted environmental effects in our quantum systems. An example is the $T_2$ dephasing process, which was also ignored in the original work~\cite{Weber2014}. The dephasing effect is undetectable by the $z$ measurements because its Lindblad, having $\hat{\sigma}_z$, commutes with the measurement backaction. Since we did not account for this dynamics, the algorithm may change the values of $\Gamma$ and $\eta$ to compensate for it, in addition to the estimation bias inherent to the model. There is also the relaxation process $T_1$; however, we find that its dynamics is insignificant and does not impact the estimation (see Appendix~\ref{app:simEst}).

\section{Conclusion}\label{sec:conC}



In this work, we have presented a hybrid online-offline algorithm for time-dependent multiparameter estimation of quantum experiments with noisy measurement data. By operating on a timescale between that of the detection rate, latency, and the parameters' drift, we can estimate hidden dynamics of interest using the SMC estimation methods augmented with the batch-averaged Kraus maps. The batch-estimation technique reduces the complexity of algorithms while still tracking any time dependent features. Our estimation algorithm consists of three main parts: (1) partitioning measurement signals into multiple batches, (2) computing batch-averaged quantum trajectories using our approximated Kraus maps, and (3) estimating the dynamics using our hybrid algorithm, which is a combination of two methods: the particle filters and the SMC sampler. 
We have also included simple overviews and derivations of the SMC methods and step-by-step details of our algorithm, hoping to encourage adoptions of the methods for any quantum experiments. 

We have implemented the algorithm to estimate multiple parameters of the two superconducting-qubit experimental datasets: the transmon's fluorescence measurement~\cite{Naghiloo2016, Naghiloo2017} and the transmon's dispersive $z$-measurement~\cite{Weber2014}. Since the parameters of interest were unknown in experiments, we have also introduced a signal reconstruction technique to benchmark our estimated parameters with the calibrated values reported in the published work~\cite{Naghiloo2016, Naghiloo2017,Weber2014}, by comparing the quality of reconstructed measurement signals with the actual raw signals from the experiments. We found that, for the fluorescence case, the four system’s parameters (i.e., Rabi frequency $\Omega/2\pi$, fluorescence decay rate $\gamma$,  measurement efficiency $\eta$, and total signal bias $\check V_{\rm off}$) estimated from our algorithm gave reconstructed signals closer to the true raw signals than the calibration parameters (see Figure~\ref{fig:sigRecon_fluo}).
Similarly, for the four parameters of the dispersive
case (i.e., Rabi frequency $\Omega/2\pi$, measurement rate $\Gamma$, measurement efficiency $\eta$, and total signal bias $\check V_{\rm off}$), our estimation gave better estimation quality (see Figure~\ref{fig:sigRecon_z}), and was even capable of finding an unexpected jump in the signal bias $\check V_{\rm off}$ during the experimental run, that the calibration method could not find (see Figure~\hyperlink{fig:allParam_zMeas}{4.d}).

However, while the SMC methods are theoretically optimal due to their foundation in Bayesian statistics, they are still impacted by the noise in the experimental data and other practical constraints. Our simple implementation of the algorithm can be improved in many aspects.
One example is a smarter choice of proposal distribution and resampling algorithm. As our choices are intentionally simple for pedagogical sakes, a more deliberate proposal will be more efficient at capturing the posterior distribution, while a more advanced resampling algorithm will deal with the degeneracy better. Another area of improvement is the dynamical model. One may improve the accuracy of the model by keeping higher-order terms in our Kraus maps or taking into account other environmental effects. Such modification can alleviate the parameter identification problem found in Section~\ref{subsec:zMeas} via better differentiation of the effects from individual parameter on the observed signals. Additionally, one may incorporate the concept of smoothing. Instead of estimating parameters $\mathbf{x}_{0:t}$ using measurement data before the estimation time $t$, i.e., $\mathbf{y}_{1:t}$, one may use the full record (which includes data after the estimation time), $\mathbf{y}_{1:n}$, to refine the estimation. This concept of smoothing has already been studied for quantum systems in Refs.~\cite{Tsang2009, Guevara2015, Chantasri2021}.
A crucial problem is the integration of the algorithm with the experimental platforms themselves. With the example of superconducting-qubit experiments, which operate in nanosecond to microsecond timescales, one may need to simplify the SMC algorithm to be able to program on Field-Programmable Gate Array (FPGA). This will also open up the possibility of augmenting the algorithm with feedback controls to further improve the estimation accuracy.


\section*{Author Contribution}
Chattamas generated and analyzed data and wrote the manuscript, Jason created the initial simulation code and reviewed the manuscript, Nattaphong contributed to Section 3 and reviewed the manuscript, Areeya and Jason supervised the project and reviewed the manuscript.

\section*{Acknowledgments}
We give our acknowledgment to Kater Murch, Mahdi Naghiloo, and Steven Weber for access to relevant experimental data. A.~C.~would like to thank Howard M. Wiseman for valuable discussions during the initial development of this idea. C.~M.~would also like to thank Nattaporn Thongphaijit for helpful discussion. This work was supported by the Program Management Unit for Human Resources and Institutional Development, Research and Innovation (Thailand) grant B05F640051.

\bibliographystyle{quantum}


\onecolumn

\appendix

\section{Full form of the signal-averaged map} \label{app:fullMap}

Here we present a first-order expansion of the signal-averaged map Eq.~\eqref{eqn:avgMap}, 
\begin{align}
    {\cal K}(\bar y_t,\mathbf{x}_t)[\rho] =\!\! \frac{1}{\ns} \sum_{j=1}^{\ns} \frac{\hat K(y_{t,j},\mathbf{x}_t) \rho \hat K( y_{t,j},\mathbf{x}_t)}{{\rm Tr}[\hat K(y_{t,j},\mathbf{x}_t) \rho \hat K(y_{t,j},\mathbf{x}_t)]}. \nonumber
\end{align}
From Eq.~\eqref{eq:evolvestate}, we only need to approximate the measurement superoperator $\mathcal{M}$. We may keep the exact form of the unitary and decoherence superoperators,
\begin{equation}\label{eq-avgmapsim}
    \mathcal{K}(\bar{y}_t,\mathbf{x}_t)[\rho]=\frac{1}{\ns}\sum_{j=1}^{\ns} \frac{\mathcal{M}(y_{t,j}, \mathbf{x}_t)[\tilde{\rho}]}{\mathrm{Tr}[\mathcal{M}(y_{t,j}, \mathbf{x}_t)[\tilde{\rho}]]}
\end{equation}
where $\tilde{\rho} \equiv \mathcal{Q}(\mathbf{x}_t)[\mathcal{U}(\mathbf{x}_t)[\rho]]$ and $\bar{y}_t = \sum_j^{\ns}y_{t,j}/\ns$. The measurement superoperator can be written as Eq.~\eqref{eqn:measureMap}, with $\hat{M}(y_t,\mathbf{x}_t)=\hat{1}-\tfrac{1}{2}\op c_{\rm o}^\dagger\op c_{\rm o}\ddt+\op c_{\rm o} y_t \ddt$. We expand Eq.~\eqref{eq-avgmapsim} up to $O(\ddt^2)$ to get
\lipsum[0]
\begin{multline}\label{eq-expandmap}
      \frac{1}{\ns} \sum_j^{\ns} \frac{\mathcal{M}(y_{t,j}, \mathbf{x}_t)[\tilde{\rho}]}{\mathrm{Tr}[\mathcal{M}(y_{t,j}, \mathbf{x}_t)[\tilde{\rho}]]} \approx \tilde{\rho} + \mathcal{Y}_t\eta\ddt(\hat{c}_\mathrm{o}\tilde{\rho}\hat{c}^\dagger_\mathrm{o}) - \frac{\eta\ddt}{2}\{\hat{c}^\dagger_\mathrm{o} \hat{c}_\mathrm{o},  \tilde{\rho}\}
      \\+ \left[\bar{y}_t\sqrt{\eta}\ddt - \mathcal{Y}_t \eta\ddt^2\mathrm{Tr}\left( \hat{c}_\mathrm{o}\tilde{\rho} + \tilde{\rho}\hat{c}^\dagger_\mathrm{o} \right) \right]\left( \hat{c}_\mathrm{o}\tilde{\rho} + \tilde{\rho}\hat{c}^\dagger_\mathrm{o} \right)
      \\
      + \left[  \mathcal{Y}_t \eta\ddt^2 \mathrm{Tr}\left( \hat{c}_\mathrm{o}\tilde{\rho} + \tilde{\rho}\hat{c}^\dagger_\mathrm{o} \right)^2 - \mathcal{Y}_t \eta\ddt^2 \mathrm{Tr}\left(\hat{c}_\mathrm{o}\tilde{\rho}\hat{c}^\dagger_\mathrm{o}\right) -\bar{y}_t \sqrt{\eta}\ddt \mathrm{Tr}\left( \hat{c}_\mathrm{o}\tilde{\rho} + \tilde{\rho}\hat{c}^\dagger_\mathrm{o} \right) + \eta \ddt \mathrm{Tr}\left(\hat{c}_\mathrm{o}\tilde{\rho}\hat{c}^\dagger_\mathrm{o}\right)\right] \tilde{\rho}
\end{multline}
where
\begin{align}\label{eq-msrecord}
    \mathcal{Y}_t \equiv \frac{1}{\ns}\sum_j^{\ns}y_{t,j}^2 = \mathrm{Var}(y_t) + \bar{y}_t^2.
\end{align}
In practice, we reduce computational cost by calculating $\mathrm{Var}(y_t)$ and $\bar{y}_t$ from the rescaled signals $v_{t,j}$ using the linear transformation of random variables.
Finally, we note that the state calculated from Eq.~\ref{eq-expandmap} must be normalized after each time step since we have neglected the contribution from the higher-order term $O(\ddt^2)$.

\section{Numerical simulation of estimation algorithm and hyperparameter search}\label{app:simEst}

This appendix discusses the estimation results from simulated data, which are used as a benchmark for the appropriate hyperparameter values. The hyperparameters of interested are discussed in Section~\ref{sec-implement}; they are the number of trajectories per batch $\ns$, the number of particles $\np$, the threshold value $\nth$, and the defensive strategy ratio $p, q$.


\subsection{Fluorescence measurement of transmon qubit}

This section shows estimation results of the simulated data from the dynamical model in Section~\ref{subsec:fluo}.

We first describe the simulation process. The system parameter values are chosen (arbitrary in the range of the typical values in experiments): the Rabi frequency $\Omega/2\pi = 0.8$, the decay rate $\gamma = 2.1$, the measurement efficiency $\eta = 0.4$, and the residual bias $v_\mathrm{off} = 0.008$. The initial Bloch-sphere coordinates are $\mathbf{r}_0 = (0.6,0,0.8)$. We generate the simulated data by evolving and measuring the system with time interval $\ddt = 0.02$. Our goal is to obtain $100 \times 10\,000$ measurement trajectories, each with $95$ data points. This dataset is similar in size to the experimental dataset~\cite{Naghiloo2016, Naghiloo2017}. 

The evolution of the state follows the Kraus map,
\begin{equation}
    \rho_t \propto \mathcal{M}(y_t, \mathbf{x}_t)[\mathcal{Q}(\mathbf{x}_t)[\mathcal{U}(\mathbf{x}_t)[\rho_{t-1}]]],
\end{equation}
given the definition of superoperators from Eq.~\eqref{eqn:fluoMap}. The associated measurement result is generated by
\begin{equation} \label{eqn:simReadOut}
    y_t = \sqrt{\eta\gamma}\,\mathrm{Tr}[ \hat{\sigma}_x\tilde{\rho}_{t-1} ] + \frac{\Delta W}{\ddt}
\end{equation}
where $\tilde{\rho}_{t-1} \equiv \mathcal{D}(\mathbf{x}_t)[\mathcal{U}(\mathbf{x}_t)[\rho_{t-1}]]$ and $\Delta W \sim \mathcal{N}(0|\ddt)$ is the Wiener increment. The Lindblad operator in this case is $\hat{c}_\mathrm{o} = \sqrt{\eta\gamma}\hat{\sigma}_-$. Afterward, we rescale the measurement results to obtain $v_t = y_t + v_{\mathrm{off},t}$ and the measurement signals $V_t = v_t\Delta V \ddt + V_\mathrm{off}$, where $V_\mathrm{off} = 125.4$ and $\Delta V = 137.3$.

We now explain the choices of hyperparameter values. We choose the common threshold value $\nth = \np/2$. To find the values of the ratio for defensive strategy $p, q$, we fix the number of trajectories per batch $\ns = 10\,000$ and the number of particles $\np = 1024$, then we test a different combination of $p, q$. We find that the differences are minimal; we decide on the $p = 0.9$ and $q = 0.1$ since it is the most efficient. 

\begin{figure*}[p]
    \centering
    \includegraphics[width=\linewidth]{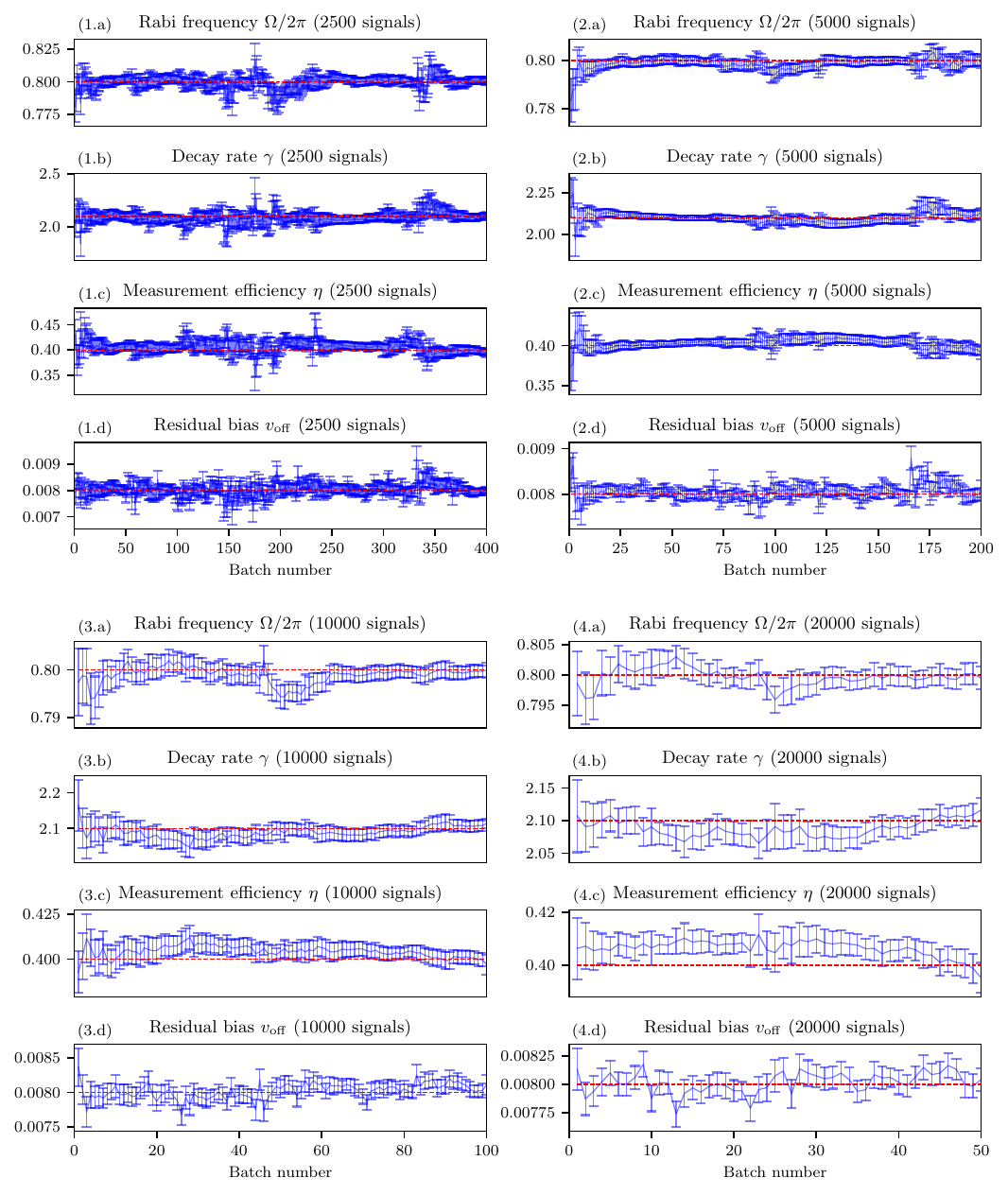}
    \caption{This figure compares the performance of the SMC algorithm at estimating the fluorescence-measurement parameters when we vary the number of trajectories per batch $\ns = \{2500, 5000, 10\,000, 20\,000\}$. We have fixed the resampling ratio $p,q = 0.1$ and the number of particles $\np= 1024$. The dash red lines are the true parameter values while the blue lines with error bars are the estimated values with their associated uncertainties. We show the residual bias $v_\mathrm{off}$ without adding the baseline scaling for ease of viewing.}
    \label{fig:sigTest_fluo}
\end{figure*}

\begin{figure*}[p]
    \centering
\includegraphics[width=\linewidth]{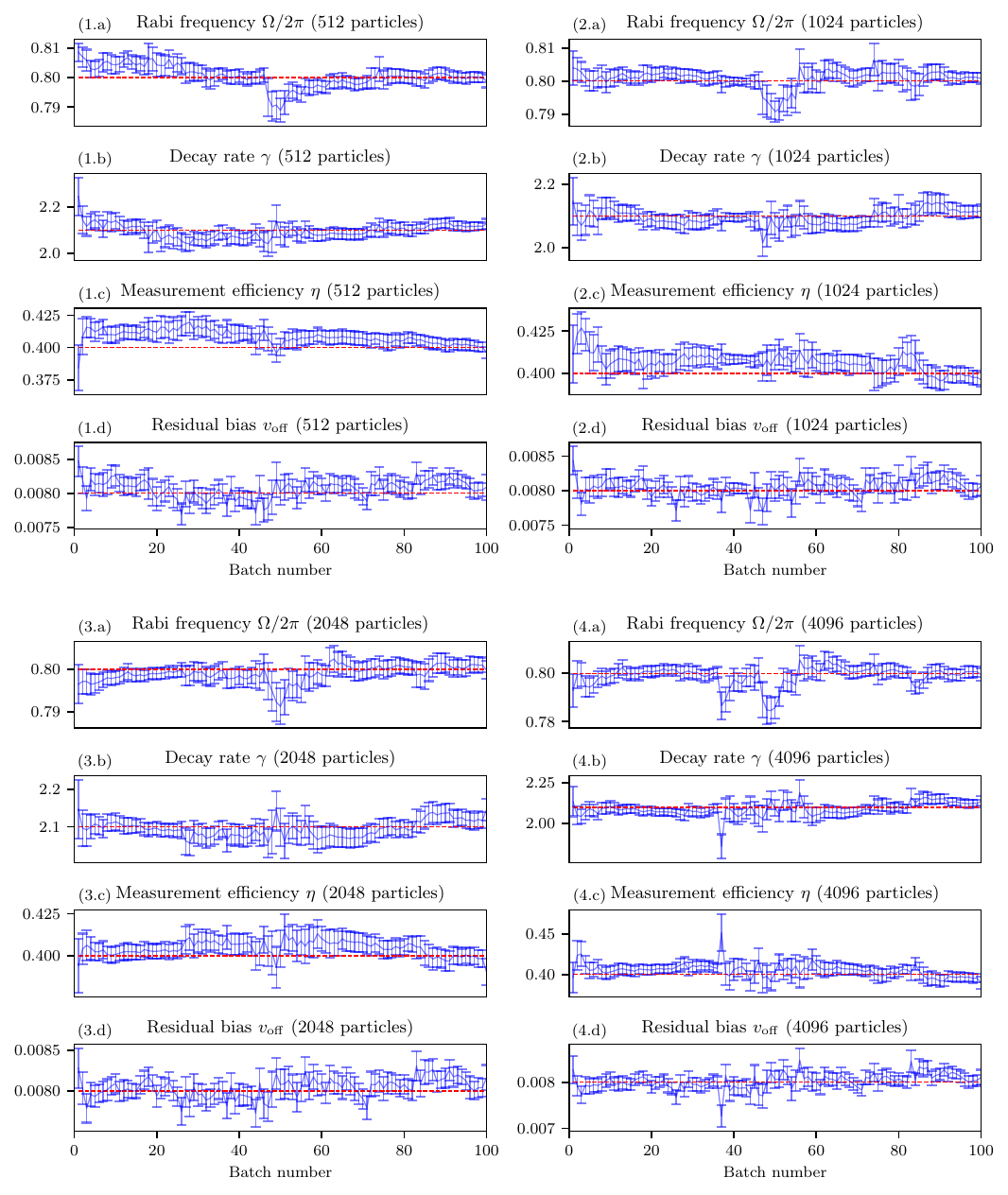}
    \caption{This figure compares the performance of the SMC algorithm at estimating the fluorescence-measurement parameters when we vary the number of particles $\np = \{512, 1024, 2048, 4096\}$. We have fixed the resampling ratio $p,q = 0.1$ and the number of trajectories per batch $\ns = 10\,000$. The dash red lines are the true parameter values while the blue lines with error bars are the estimated values with their associated uncertainties. We show the residual bias $v_\mathrm{off}$ without adding the baseline scaling for ease of viewing.}
    \label{fig:parTest_fluo}
\end{figure*}

The more significant hyperparameters are $\ns$ and $\np$. To find $\ns$, we fix $\np = 1024$, and vary the value of $\ns = \{2500,5000,10\,000,20\,000\}$. Figure~\ref{fig:sigTest_fluo} shows the estimation results of the test. It is clear that large $\ns$ compensate better signal-to-noise ratio with the reduced the number of batches. We have chosen $\ns = 10\, 000$ because it gives us a balance between the accuracy and the computational cost.
The same method is used to find $\np$; we fix $\ns = 10\,000$ and vary $\np = \{512, 1024, 2048, 4096\}$. Figure~\ref{fig:parTest_fluo} shows the estimation results of the test. Higher numbers of particles provide better accuracy at the cost of increased computational times. In the end, we choose $\np = 1024$ for a balance of speed and accuracy.

In summary, we choose the ratio $p = 0.9$ and $q=0.1$, $\nth = \np/2$, $\ns = 10\,000$, and $\np = 1024$ for the fluorescence setting.

On a final note, consider the case of faulty data-cleansing. The rescaled signal $v_t$ comes from the raw data $V_t$,
\begin{equation}
    v_t = \frac{V_t - V_\mathrm{off}}{\Delta V \ddt}.
\end{equation}
There can be two points of failures: the baseline voltage $V_\mathrm{off}$ and the scaling factor $\Delta V$. If the two parameters have incorrect values, the resulting measurement result $y_t$ will be wrong, affecting the estimation algorithm. An inaccurate baseline voltage $V_\mathrm{off}$ is less problematic since any difference can be absorbed into the residual bias $v_\mathrm{off}$. However, an inaccurate scaling factor $\Delta V$ impacts the measurement result, i.e., $y'_t = y_t/\alpha$, where $\alpha$ is the mismatched ratio. From Eq.~\eqref{eqn:simReadOut}, $\alpha$ can be absorbed into the product $\eta \gamma$ and the Wiener increment $\Delta W' = \Delta W/\alpha \sim \mathcal{N}(0| \ddt / \alpha^2)$. This likely impacts the estimation results of the system parameters. We have tested such effect by applying the scaling factor $\alpha = \{0.5, 2.0\}$ to the measurement signals and found the impacted parameters to be measurement efficiency $\eta$ and decay rate $\gamma$. The algorithm compensates for the wrong scaling factor by over- or under-estimating a product, $\eta\, \gamma$, impacting the estimation of $\eta$ and $\gamma$ individually.

\subsection{Dispersive $z$-measurement of transmon qubit}

This section shows the estimation results of the simulated data from the dynamical model in Section~\ref{subsec:zMeas}.

The chosen system parameter values are the Rabi frequency $\Omega/2\pi = 1.2$, measurement rate $\Gamma = 4.0$, measurement efficiency $\eta = 0.4$, relaxation constant $1/T_1 = 0.025$, and residual bias $v_\mathrm{off} = 0.8$. The initial Bloch-sphere coordinates are $\mathbf{r}_0 = (0.88, 0, 0)$. Our goal is to obtain $200 \times 1000$ trajectories, each with $90$ data points and time resolution $\ddt = 0.016$. All parameter values are motivated by those in Ref.~\cite{Weber2014}.

The evolution of the state follows the Kraus map,
\begin{equation}
    \rho_t \propto \mathcal{M}(y_t, \mathbf{x}_t)[\mathcal{Q}_2(\mathbf{x}_t)[\mathcal{Q}(\mathbf{x}_t)[\mathcal{U}(\mathbf{x}_t)[\rho_{t-1}]]]],
\end{equation}
given the definition of superoperators from Eq.~\eqref{eqn:diszMap}, with an additional decoherence from $T_1$ relaxation $\mathcal{Q}_2$.  We note that the relaxation time is usually much longer than the experimental duration and the measurement period $T_1 \gg 1/\Gamma$, the impact of $\mathcal{Q}_2$ is minimal to the evolution. Our result shows that for $1/T_1 \ll 1$, the algorithm cannot precisely estimate $1/T_1$. This is why we have neglected the estimation of $T_1$ using the experimental data. For the numerical simulation, the generated measurement readouts is in the same form as in Eq.~\eqref{eqn:simReadOut} with the Lindblad operator $\hat{c}_\mathrm{o} = \sqrt{\eta\Gamma/2}\hat{\sigma}_z$. Afterward, we rescale the measurement readouts to obtain $v_t = y_t + v_{\mathrm{off}}$ and the pre-processed signal $V_t = v_t \Delta V + V_\mathrm{off}$ where $V_\mathrm{off} = -1.52$ and $\Delta V = 2.055$.


Similar to the fluorescence setting, the choice of $\nth = \np/2$ is the common choice. By fixing the number of trajectories per batch $\ns=2000$ and the number of particles $\np = 2048$, we may vary the ratio $p,q$. The chosen ratios are $p=0.9$ and $q=0.1$ as they are the most efficient.

\begin{figure*}[p]
    \centering
    \includegraphics[width=\linewidth]{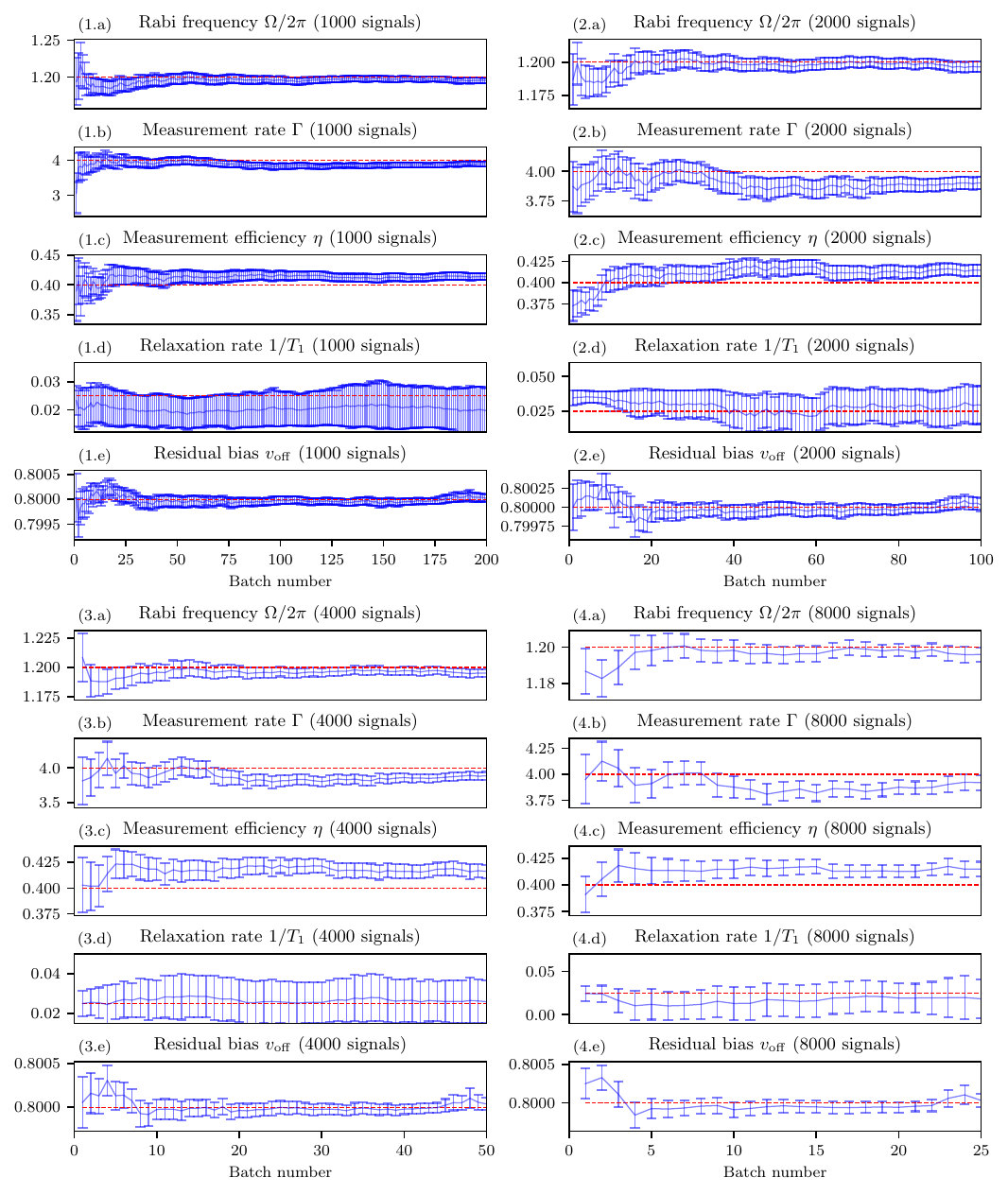}
    \caption{This figure shows the performance of the SMC algorithm for estimating $z$-measurement parameters when we vary the number of trajectories per batch $\ns = \{1000, 2000, 4000, 8000\}$. We have fixed the resampling ratio $p,q = 0.1$ and the number of particles $\np = 2048$. The dash red lines are the true parameter values while the blue lines with error bars are the estimated values with their associated uncertainties. We show the residual bias $v_\mathrm{off}$ without adding the baseline scaling for ease of viewing.}
    \label{fig:sigTest_z}
\end{figure*}

\begin{figure*}[p]
    \centering
    \includegraphics[width=\linewidth]{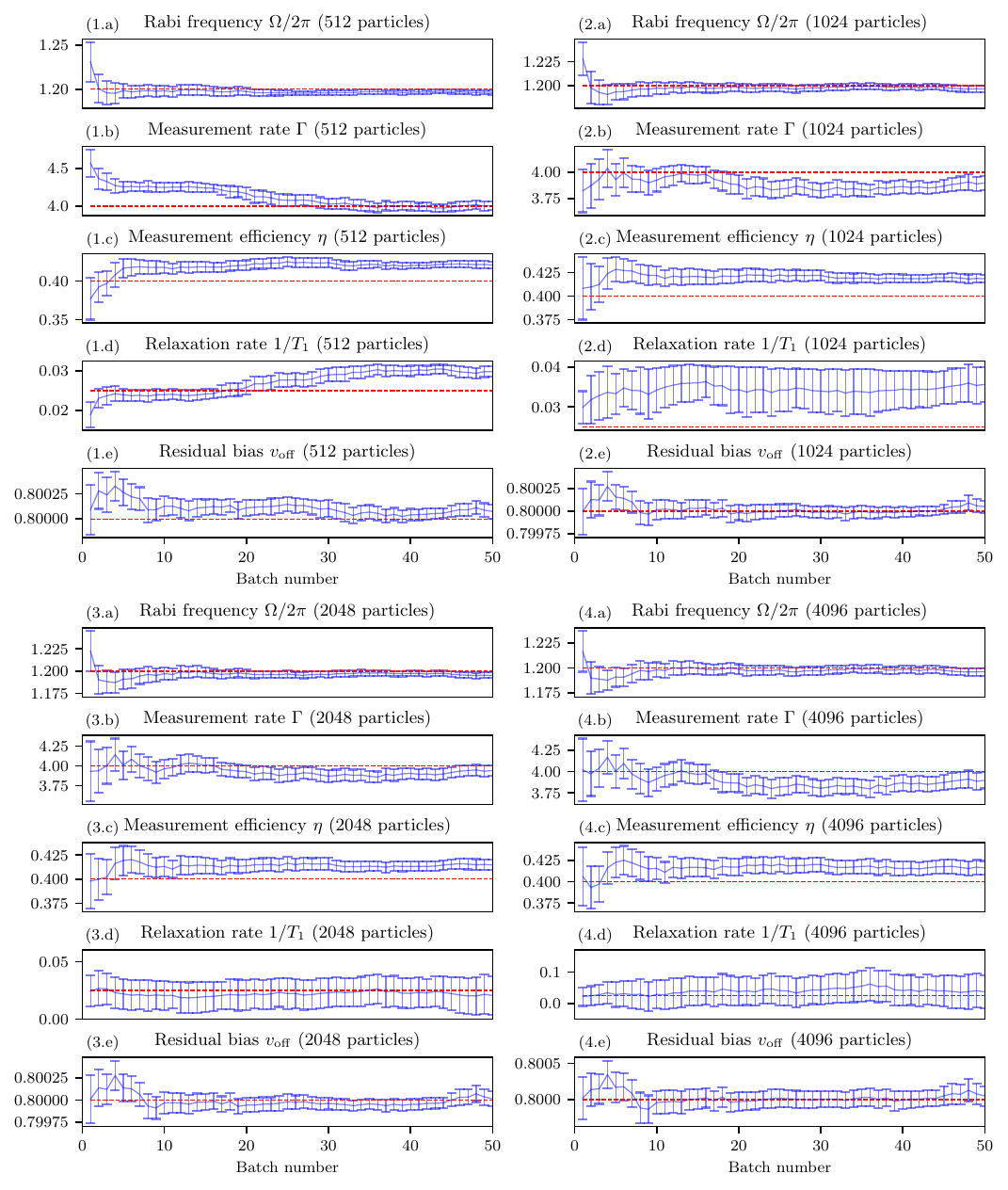}
    \caption{This figure shows the performance of the SMC algorithm for estimating $z$-measurement parameters when we vary the number of particles $\np = \{512, 1024, 2048, 4096\}$. We have fixed the resampling ratio $p,q = 0.1$ and the number of trajectories per batch $\ns = 4000$. The dash red lines are the true parameter values while the blue lines with error bars are the estimated values with their associated uncertainties. We show the residual bias $v_\mathrm{off}$ without adding the baseline scaling for ease of viewing.}
    \label{fig:parTest_z}
\end{figure*}

With the method analogous to the previous subsection, we fix $\np = 2048$ and vary the number of trajectories per batch $\ns = \{1000, 2000, 4000, 8000\}$. The best value seems to be $\ns = 4000$ as it gives an accurate estimation while still maintaining a fair amount of batches. Figure~\ref{fig:sigTest_z} shows the results. The estimation processes for $\ns = 4000$ and $\ns = 8000$ are similar, showing that $\ns = 4000$ is accurate enough. We note that the estimation for the relaxation rate $1/T_1$ is highly uncertain as the effect is negligible for our dynamics. Lastly, we show the estimation results when fixing $N_s = 4000$ and varying the number of particles $\np = \{512, 1024, 2048, 4096\}$ in Figure~\ref{fig:parTest_z}. The best value is chosen to be $\np = 2048$.

In summary, we have the threshold value $\nth = \np/2$, the resampling ratio $p=0.9$ and $q=0.1$, the number of trajectories per batch $\ns = 4000$, and the number of particles $\np = 2048$ for the dispersive $z$-measurement setting.

\end{document}